\documentclass[aps,pra,twocolumn,superscriptaddress,floatfix,footinbib]{revtex4}
\usepackage[english]{babel}
\usepackage{graphicx}   
\usepackage{bm}         
\usepackage{mathtools}
\usepackage[bbgreekl]{mathbbol}  
\usepackage{amsmath}  
\usepackage{amssymb}
\usepackage{lineno}
\usepackage{xcolor}
\DeclareSymbolFontAlphabet{\amsmathbb}{AMSb}
\usepackage{comment}
\usepackage{footnote}
\usepackage[normalem]{ulem}
\usepackage{lineno}

\usepackage{hyperref}

\usepackage{subcaption}
\newcommand{    \e  }{\mathrm{e}}   
\newcommand{    \op }[1]{\ensuremath{\hat{\rm{#1}}}}    
\newcommand{    \figref }[1]{Fig.~\ref{#1}}

\newcommand{    \secref }[1]{sec.~\ref{#1}}
\newcommand{    \Eqref      }[1]    {Eq.~\eqref{#1}}
\newcommand{    \mx }{\ensuremath{{\rm max}}}
\newcommand{    \SE}{Schr\"odinger equation}
\begin{document}
\definecolor{darkblue}{rgb}{0, 0, 0.75}
\definecolor{darkgreen}{rgb}{0, 0.5, 0}
\def\paul#1{{\color{orange}#1}}
\def\paulno#1{{\color{orange}\sout{#1}}}
\def\comment#1{{\color{darkblue}#1}}
\def\discuss#1{{\color{darkgreen}(#1)}}
    \title{Confined few-particle systems beyond mean-field theory adopting 
    Gaussian-type orbitals and Morse interparticle interaction}
    \author{Matee ur Rehman}
    \author{Paul Winter}
    \altaffiliation[Present address: ]{
        Institut f\"ur Theoretische Physik, 
        Leibniz Universit\"at Hannover,
        Appelstra\ss e 2, 30167 Hannover, Germany}
    \affiliation{
        AG Moderne Optik,
        Institut f\"ur Physik,
        Humboldt-Universit\"at zu Berlin,
        Newtonstr.~15, D\,-\,12\,489 Berlin, Germany}
    \author{Fabio Revuelta}
    \affiliation{
        Grupo de Sistemas Complejos, 
    	Escuela T\'ecnica Superior de Ingenier\'ia Agron\'omica,
    	Alimentaria y de Biosistemas,
    	Universidad Polit\'ecnica de Madrid,
    	Avenida Puerta de Hierro 2-4, 28040 Madrid, Spain}
    \author{Alejandro Saenz}
    \affiliation{
        AG Moderne Optik,
        Institut f\"ur Physik,
        Humboldt-Universit\"at zu Berlin,
        Newtonstr.~15, D\,-\,12\,489 Berlin, Germany}
    \date{\today}
%
%
%
%
%
%
\begin{abstract}
\noindent%
Recent advancements in optical tweezers enable the trapping 
of arbitrary numbers of neutral atoms and molecules, even 
arrays of tweezers with variable geometry can be realized. 
These fascinating breakthroughs require 
novel full-dimensional beyond mean-field treatments
for systems with more than two confined particles spread 
over traps that are arranged arbitrarily in space. 
In this work, the suitability of a quantum-chemistry 
inspired approach adopting Cartesian Gaussians as basis 
functions is investigated. For this purpose,
the six-dimensional integrals 
associated with a
realistic atom-atom interaction
described by a Morse model 
potential were implemented.
The performance, correctness 
and efficiency of the implementation
is assessed by comparing full 
configuration-interaction calculations (exact diagonalizations)
for two atoms in an isotropic harmonic trap 
with quasi-exact reference results.
\end{abstract}
%
%
%
%
%
    \maketitle    
%
%
%
%
%
%
%
%
\section{Introduction}\label{sec:intro}
The experimental development of depositing single atoms into optical tweezer 
arrays of in principle arbitrary shape has emerged as a powerful 
platform for quantum simulation and 
computation \cite{cold:spar22,cold:paus23,cold:bloc23}, quantum information 
processing \cite{cold:levi19}, and 
high-precision atomic clocks \cite{cold:youn20}. 
This fascinating tool
has now been expanded with the development of trapping molecules
\cite{cold:kauf21, cold:yicha21}.
Furthermore, it became possible to move tweezers or tweezer arrays 
relative to each other~\cite{cold:ti20}, allowing also for merging 
them, e.\,g., in order to replicate molecular configurations out 
of individual atoms~\cite{cold:rutt23,cold:rutt24}. An accurate theoretical description 
of these systems
strongly relies on adequately taking
into account the influence of the, possibly multi-centered, potential of 
the tweezer (arrays), especially if the spatial confinement length due to 
the trap becomes comparable to the effective 
range of the particle-particle interaction. For example, the positions 
of magnetic Feshbach resonances
\cite{cold:chin10,aies:ayma78},
that are often adopted for manipulating the interatomic interaction may 
shift due to the external trap potential~\cite{cold:schn09}.
The anharmonicity of the trap potential as well as different masses 
or polarizabilities of the particles in case of different atomic species 
(or the same atom being in different internal states) can 
even lead to the so-called inelastic confinement-induced resonances
\cite{cold:sala12,cold:hall10b,cold:cape23,cold:vali11,cold:mele11}.
These resonances can be 
used for the coherent formation of molecules \cite{cold:sala13,cold:sala16a}, 
but they can also be a nuisance in quantum-simulation experiments,
especially if their occurrence is not expected.
Furthermore, it is also of interest to 
study the influence of the confining potentials of a tweezer array 
on, e.\,g., exotic bound states like 
the Efimov trimers~\cite{Efimov1970, Efimov1970a, cold:esry99a,cold:wern06,
cold:ferl09,cold:naid17,
cold:higg22,cold:boug23}. 

In this work,
a quantum-chemistry inspired 
description of
ultracold atoms trapped in tweezer arrays in which 
the potential wells are arbitrarily distributed in space 
is discussed. Within this
approach, Cartesian or spherical 
Gaussians, preferentially but not necessarily positioned at 
the centers of the tweezer potential wells, are adopted as  
single-particle basis functions. Many-particle wave functions 
are expressed as linear combinations of properly symmetrized 
(Bosonic atoms) or antisymmetrised (Fermionic atoms) 
one-particle tensor-product states (configurations). As in 
electronic-structure calculations, the configurations may be 
formed adopting one-particle wave functions from a mean-field 
(or density-functional) calculation. Different choices of the 
configurations adopted in solving the many-particle problem 
allow for compromising between the required computational resources 
and the precision that can be achieved. 

As a first step toward the goal of implementing a general 
approach for ultracold atoms trapped in a tweezer array, 
an efficient evaluation of the two-particle integrals 
involving Gaussian basis functions was 
formulated, implemented, and tested for the case of a 
realistic model of interatomic interaction:
the Morse potential. 
Furthermore, the adequacy and convergence behavior of Gaussians 
for describing trapped ultracold atoms in the regime in which 
the interatomic interaction length becomes comparable to the 
trap length was investigated for the simple example of 
two interacting ultracold atoms in an isotropic 
harmonic single-well potential, since quasi-exact reference 
results can be obtained for this case using, e.\,g., the 
numerical approach described in Ref.~\cite{cold:gris09}.

This paper is organized as follows.
Section~\ref{sec:uca_in_ot} describes the problem of interest, 
the existing numerical approaches, and their limitations.
Furthermore, it includes a motivation 
for the proposed approach,
and discusses the challenges that need to be overcome.
Section~\ref{sec:method} introduces the method
with a particular emphasis on the evaluation of the 
two-particle integrals for the Morse interaction potential 
considered in this work. 
Section~\ref{sec:details} outlines the computational details, 
i.\,e., the parameters governing the trap potential, the 
interatomic Morse potential, and the basis-set specifications. 
Section~\ref{sec:results} presents and discusses the results,
i.\,e., basis-set convergence and a comparison of the energy 
spectrum and eigenfunctions with reference data. 
Finally, in Sec.~\ref{sec:conclusions}, an outlook and 
a discussion on the future research directions and applications 
is provided.

\section{Ultracold atoms in tweezer arrays}
\label{sec:uca_in_ot}
The precise and thus fully correlated, three-dimensional theoretical description 
of trapped atoms and
molecules adopting realistic particle-particle interaction potentials is very challenging. 
This task is particularly demanding
if the trap dimension is similar to the effective atom-atom 
interaction length, since then both, the trap potential 
and the interparticle interaction potential,
need to be treated 
simultaneously and on a comparable level of accuracy. 
The computational demands become particularly high in 
situations where atomic or molecular wave functions are 
delocalized across multiple sites in a tweezer array.
These challenges are further amplified when the array 
consists of wells that are arbitrarily distributed in space,
and potentially vary in shape, i.\,e., when the wells lack 
the symmetric arrangement characteristic of an optical lattice.

As a consequence, many well-established approaches developed for 
accurately describing trapped ultracold atoms or molecules cannot
be straightforwardly, or even at all, extended to this 
more intriguing situation.
For example, a reduction of the 
dimensionality as is, e.\,g., adopted in
the Refs.~\cite{cold:boug22,cold:boug19,cold:boug21},
that reduces the Hilbert space of $N$ particles to the one 
describing $N$ (one dimensional) or $2N$ (two dimensional) 
instead of $3N$ degrees of freedom, would be limited with 
respect to the type of tweezer potential and
the geometry of the tweezer array that can be described, 
since then both need to be correspondingly (quasi-)one 
or two dimensional. Moreover, even if one considers 
three quasi-one dimensional tweezer potentials, a one-dimensional 
treatment is limited to the case
where all potential wells 
are aligned along one line. In a non-linear 
arrangement, even a two-dimensional treatment would be insufficient, 
if the one-dimensional potentials of the three tweezers are 
not properly aligned with respect to the plane spanning the triangle. 
There are, of course, also experimental situations in which the 
full dimensionality plays a central role as, e.\,g., 
in the one described in Ref.~\cite{cold:cape23}.

The second approximation that is often adopted in the treatment of 
ultracold quantum gases concerns the particle-particle interaction. 
Even in the case of an isotropic interaction, 
the full three-dimensional treatment of two or more particles based on basis 
functions requires the calculation of six-dimensional 
two-particle integrals.
(In fact even higher dimensional ones 
if geminals, i\,e., two-particle 
basis functions are adopted.)
What is more,
the number of these integrals increases with the 4$^{\rm th}$ power 
of the number of single-particle basis 
functions in a naive approach, but still exponentially, 
even if numerous tricks are applied. The most popular 
approximation is to describe the 
interparticle interaction via a pseudopotential 
given by a (regularized) $\delta$ function. Undoubtedly, 
this leads to a drastic simplification of 
the two-particle integrals,
reducing the dimension from~6 to~3.
However, in the presence of an external (trap) potential and in a full 
three-dimensional treatment, the (regularized) $\delta$-function 
pseudopotential fails to converge in beyond-mean-field 
approaches like exact diagonalization \cite{cold:esry99}.
Therefore, a renormalization is required
(see Ref. \cite{cold:brau25} 
and references therein),
which, in turn, massively 
complicates the treatment. In fact, already within the 
mean-field description,
the standard $\delta$ pseudopotential 
employing the trap-free scattering length does not lead to 
an accurate description. Instead, an energy-dependent 
scattering length \cite{cold:blum02} is usually adopted that is, however, 
only available after a full treatment of the correct interparticle 
interaction {\it together} with the confining trap potential. 

Other simplified pseudopotentials are the harmonic or the Gaussian 
potentials, as they are, e.\,g., adopted in 
the Refs.~\cite{cold:arms11}
and~\cite{cold:higg22,cold:yin14,cold:jesz18,cold:jesz19},  
respectively. As in the case of the $\delta$ pseudopotential,
they do not reproduce at all the shape
of the physical particle-particle
interaction potential, but can provide wave functions with the 
proper behavior at large interparticle separations. In contrast to the 
case of the $\delta$ potential, the two-particle integrals remain 
six dimensional, but are often much easier to evaluate, especially 
if (Cartesian) Gaussian basis functions~\cite{math:boys50} or 
Gaussian geminals~\cite{cold:raks12,math:barc17} are used.  

In order to
accurately describe two ultracold atoms in a 
multi-well trapping potential like a (finite) optical lattice, 
a configuration-interaction (CI) approach was introduced in 
the Refs.~\cite{cold:gris09,cold:gris11}.
The Hamiltonian is partitioned into three terms: 
one depending only on the coordinate of the center-of-mass motion,
one depending only on the one of the relative motion,
and one describing the coupling in between.
The configurations are then built 
as products of the center-of-mass and the relative-motion wave functions. 
This approach allows for a very efficient treatment of the 
inter-particle interaction, since it enters only the 
relative-motion Hamiltonian. 
Expressing the radial part of the relative-motion wave functions 
in $B$ splines and the angular part in spherical harmonics, 
the inter-particle interaction integrals become effectively one-dimensional 
in the case of an isotropic interaction potential. This allows 
for the integral calculation with minimal overhead, even if numerical 
Born-Oppenheimer interaction potentials are adopted. In fact, 
even anisotropic dipole-dipole interactions can be considered, 
if the dipoles are aligned~\cite{cold:schu15,cold:schu16}. 
Unfortunately, this approach cannot 
be directly extended to more than two particles. The most analogous extension
would be one adopting Jacobi coordinates,
see, e.\,g., Refs.~\cite{cold:dail10,cold:rotu10,cold:gree17,gen:hiya18}.
However, the number of coordinates 
and thus the computational efforts increase drastically with 
the number of particles. Furthermore, the incorporation of 
a multi-centered trap potential appears non-trivial. 
Similarly, so far the use of hyperspherical 
coordinates, see, e.\,g., 
Refs.~\cite{cold:von09,cold:sze18,cold:higg22}, 
seems to be limited to at most 
isotropic (harmonic) trap potentials and an extension to the 
multi-center case appears to be very challenging and 
far from being straightforward.

Recall that, in the long run,
the problem of interest is a system of  
possibly many interacting ultracold atoms (or molecules) 
that are trapped in a multi-centered optical tweezer array 
whose potential wells may be arbitrarily distributed in space.  
Noteworthy, this closely resembles the electronic-structure 
problem in molecules \cite{cold:sala17}.
The potential wells formed 
by the optical traps correspond
(within the Born-Oppenheimer approximation)
to the nuclear (or ionic core) potentials in 
molecules, while the interacting ultracold atoms (or molecules) 
correspond to
the electrons.
In view of the impressive efficiency of the 
various quantum chemistry packages in solving the 
electronic-structure problem of molecules,
it appears natural to adopt the same concepts 
also to the challenging task of describing ultracold particles 
in tweezer arrays. Most of the quantum-chemistry approaches 
adopt Gaussian basis functions, here called Gaussian-type 
orbitals (GTOs), for expanding the electronic wave functions, 
despite the fact that Gaussians neither possess the proper 
shape close to the nuclei nor at large distances. However, 
in contrast to the proper hydrogen-like (exponential) 
basis functions, called Slater-type orbitals, Cartesian GTOs 
allow for extremely efficient calculations of the multi-centered 
two-electron integrals. Although Cartesian GTOs require a 
larger number of basis functions than Slater-type orbitals
for achieving the same accuracy,
this drawback is typically 
overcompensated by their efficiency in 
the calculation of the interaction integrals,
which is often the major bottleneck in electronic structure 
computations. 

In the spirit of the quantum-chemistry approaches, GTOs placed at 
the centers of the various potential wells of the tweezer array 
may be used as basis functions. For a more accurate description, 
additional Gaussians may be positioned in between the wells, as 
is done in electronic-structure calculations, e.\,g., for a 
better description of chemical bonds. Within mean-field (or 
density-functional) theory, the time-independent Schr\"odinger 
equation is solved in order to obtain single-atom wave functions 
(corresponding to electronic orbitals in molecules). They are 
a good starting point for including (dynamic) correlation by 
CI, coupled-cluster theory, and other 
approaches like multi-configuration self-consistent field methods, 
i.\,e., for obtaining correlated many-atom wave functions. 
While approaches adopting explicitly correlated basis functions 
are known to be much more efficient in coping with correlation, 
they tend to become quickly intractable for an increasing number 
of particles. 

Compared to the electronic-structure problem, the use of GTOs 
for trapped ultracold atoms is, in fact, even more appropriate 
than for electrons in molecules, since the trap potentials for 
neutral atoms are usually, in first-order approximation, harmonic. 
Note, however, that the description of a tweezer array as 
a set of perfectly harmonic (or polynomial) potentials is not 
suitable, as these lead to infinitely high potential barriers 
or wells. On the other hand, 
sets of Gaussian wells are 
very flexible and suitable for describing 
tweezer-array potentials, and, as a consequence, 
lead to straightforward integrals describing the attractive 
atom-trap interaction.  

There is, however, also a number of challenges when trying to 
transfer existing 
electronic-structure codes to ultracold atoms. 
First of all, unlike electrons, which are always spin-1/2 Fermions,  
ultracold atoms can be Fermions with spins different from 1/2, 
or they can also be Bosons with integer spins. Furthermore, 
they do not need to be all identical atoms. 
Most importantly, the (effective) interaction potential between 
two ultracold atoms is usually only known from molecular 
Born-Oppenheimer calculations, spectroscopy, or a combination 
of them. These potentials depend on the considered pair of 
atoms and are certainly much more involved
than the analytically given pure Coulombic repulsion present 
among electrons, since they are only numerically provided
as a function of the interatomic separation. Thus,
one of the key questions is whether the legion of two-particle 
integrals involving up to four different, possibly multi-centered,
GTOs can be evaluated with sufficient precision and efficiency for
realistic interatomic interaction potentials. 
This challenge arises because the sophisticated and 
ingenious algorithms designed for the efficient computation 
in the case of pure
Coulombic interactions, which lie at the 
core of quantum-chemistry codes,
cannot be directly (or at all) applied. 

In this work, the integrals involving GTOs have been implemented for 
an interatomic interaction 
described by a realistic Morse model potential.
The approach is motivated by Refs.~\cite{math:silk15a,gen:balc17}
which detailed how the six-dimensional two-particle integrals involving 
the product of up to four different (possibly 
multi-centered) GTOs and an isotropic function that depends on 
the relative coordinate can be reformulated as a multiple sum 
over one-dimensional master integrals.  

In comparison to the electronic-structure problem, where the 
inter-particle interaction is purely repulsive, most of the 
relevant interatomic interaction potentials are at intermediate 
interatomic separation attractive, supporting (often numerous) 
molecular (diatomic) bound states.
This leads to the question 
whether GTOs centered at tweezer wells are suitable for describing 
trapped ultracold atoms in a regime where the (effective) 
interaction between the atoms (that can be attractive or 
repulsive) becomes comparable to the trap potential. In contrast 
to most quantum-chemistry calculations that concentrate on the 
electronic ground state or possibly some low-lying excited 
states, in the case of trapped ultracold atoms the interest 
often involves a series of states, e.\,g., in studies of 
inelastic confinement-induced resonances~\cite{cold:sala12,cold:sala16a,cold:cape23}.
Similarly, investigating relaxation dynamics and thermalization in many-body 
quantum systems necessitates the computation of excited eigenfunctions. 
The behavior of these systems can vary significantly depending on whether 
the eigenfunction thermalization hypothesis 
is satisfied \cite{cold:deut91,cold:sred94,cold:shir21}. 
Notably, deviations from the previous hypothesis 
--- such as those arising from quantum many-body scarring 
\cite{cold:moud18,cold:turn18,cold:serb21} ---
lead to atypical dynamical behavior. 
Eigenfunctions of excited states are also central to the phenomenon of 
single-particle scarring, which is crucial for understanding the unexpected 
localization of wavefunctions along unstable periodic orbits \cite{cold:revu12}.

In order to investigate the suitability of GTOs in the context 
of trapped ultracold atoms, this work considers as a first 
step only the case of two atoms interacting via a Morse potential 
and
confined in
an isotropic harmonic trap.
For this situation,
quasi-exact reference data can be obtained, allowing for a 
well-defined check of both the correctness of the implementation 
as well as the convergence behavior of the GTOs toward the 
correct result. The GTOs optimized for a single-well 
potential should serve as a good starting point for future 
multi-well calculations. This is very similar to quantum chemistry, where
GTOs optimized for the description of 
atoms are often adopted in molecular calculations or used as a starting point 
for further optimization.
%
%
%
%
%
%
%
%
\section{Method}\label{sec:method}
\subsection{Configuration-interaction with Gaussians}
The stationary \SE\ in absolute coordinates for
{\bf $N$} identical particles trapped in a 
three-dimensional
potential is 
    \begin{align}\label{eq:system:schrodinger}
   	\left[\sum_{i=1}^{N}\left(-\frac{\hbar^2}{2m_{\rm i}} \nabla^2_{\mathbf r_i}  
    + \op{V}_{{\rm
    trap},i}(\mathbf r_i) + \sum_{j>i}^N{\op{U}}({ r_{ij}})  \right) \right] \Psi
   	= 
   	E \, \Psi,
   \end{align}
where $\Psi=\Psi(\mathbf r_{1},...,\mathbf r_{N})$ is the wave function 
for the {\bf $N$} particles. The first term in the sum is the 
kinetic-energy 
operator 
of the $i$-th particle and 
$\op{V}_{{\rm trap},i} (\mathbf r_i)$ is the external (tweezer-array) 
trapping potential. Finally, $\op{U}({ r_{ij}})$ is the interatomic-interaction 
potential that is 
assumed to be isotropic and thus to depend 
solely on the interatomic distance ${ r_{ij}}=|\mathbf r_i -\mathbf r_j|$. 

Within the CI ansatz \cite{gen:szab96}, also 
known as exact diagonalization, the wave function $\Psi$ is expressed as 
a linear combination of $N_{\rm CF}$ configurations $\Phi$
\begin{equation}\label{eq:method:AWF}
   \Psi(\mathbf r_{1},...,\mathbf r_{N}) = \sum_{j=1}^{N_{\rm CF}} c_j \: 
           \Phi_j (\mathbf r_{1},...,\mathbf r_{N})\; .
\end{equation}
These configurations,  
\begin{equation}\label{eq:method:WF}
	\Phi_j (\mathbf r_{1},...,\mathbf r_{N}) = \op{S}^{(\pm)}_{N} \: 
          \prod\limits_{\tau=1}^{N} \: \phi_\tau(\mathbf r_\tau) \ , 
\end{equation}
are (anti-)symmetrized ($-$/$+$) products of 
one-particle wave functions
$\phi$, where
$\op{S}^{(\pm)}_N$ is the (anti-)symmetrization operator, which 
may include an appropriate normalization factor.
If all configurations $\Phi$ 
that can be constructed from a given set of one-particle 
wave functions are included, the approach is known as full CI. 
Full CI calculations are computationally extremely expensive 
and, in practice,
restricted to a very small number of particles 
and one-particle wave functions. Therefore, 
most of the CI calculations include only a (very) limited 
number of configurations instead of the complete set. 
In order to still obtain good results, 
both the selection of the configurations $\Phi$ as well 
as the quality of the 
adopted one-particle wave functions $\phi$ is of crucial 
importance. 

The traditionally most common choice
consists in building
a reference configuration from the mean-field 
(Hartree-Fock) orbitals that are occupied in the ground state. 
Then, all the configurations that can 
be obtained by transferring one (or two) electron(s) from 
the reference configuration to an unoccupied orbital
are added. This is known as CI singles (or CI 
singles and doubles). The optimization 
of the choice of configurations remains 
an active subject of investigation.
For the simple two-particle example  
considered in this exploratory work,
full CI is adopted. Since the mean-field 
orbitals and the basis functions are related 
by a unitary transformation, identical results are obtained 
in a full CI, if the configurations are either built from 
mean-field orbitals or directly from the basis functions.
Thus, no mean-field orbitals are calculated in this work.

Usually, multi-center integrals involving GTOs are most 
suitably calculated in Cartesian coordinates. Therefore, 
Cartesian GTOs, defined as 
\begin{equation} \label{eq:method:GTOs} 
   \phi_\tau(\mathbf r) 
      = \mathcal{N}_{\tau} 
       (x-A_{x})^{i}\, (y-A_{y})^{k}\, (z-A_{z})^{m}
        \e^{-\tau(\mathbf{r}-\mathbf{A})^2},
\end{equation}
are often adopted in quantum-chemistry calculations. 
They are thus characterized by the set of parameters 
$i,k,m, \tau$, and $\mathbf A$, with
$i,k,m\in \mathbb N_0$, the width of the orbital 
$\tau \in \mathbb R^+$, the center of the 
Gaussian $\mathbf A = (A_x,A_y,A_z)\in \mathbb R^3$, 
and the normalization factor 
\begin{equation}\label{eq:method:NF}
  \mathcal{N}_{\tau} = \left(\frac{2\tau}{\pi}\right)^{\frac{3}{4}}
\sqrt{\frac{(4\tau)^{i+k+m}}{(2i-1)!! (2k-1)!! (2m-1)!!}} \;.
\end{equation}
The number of Cartesian Gaussians for a given value 
of $\sigma:=i+k+m$
is $(\sigma+1)(\sigma+2)/2$. 
Only for $\sigma=0$ or~$1$ there is (in the isotropic case) 
a one-to-one correspondence with the quantum number of 
angular momentum $\ell$, i.\,e., the
single
Cartesian 
GTO with $\sigma=0$ corresponds to an s-type 
orbital ($\ell=0$) and the three GTOs with $\sigma=1$ 
correspond to the p-type orbitals with $\ell=1$ 
(p$_x$, p$_y$, and p$_z$). However, there are 6 
Cartesian GTOs for $\sigma=2$ (10 for 
$\sigma=3$, etc.) 
that can be shown to correspond to 5 d-type orbitals 
with a quantum number of angular momentum $\ell=2$ and~1
additional s-type orbital with $\ell=0$ (7 f-type 
orbitals with $\ell=3$ and 3 p-type orbitals with 
$\ell=1$, etc.). It is common practice to name the  
Cartesian GTOs according to the value of $\sigma$, e.\,g., 
the $\sigma=2$ ones are called d-type GTOs, despite the fact 
that they contain also an $\ell=0$ orbital. 
Removing the linear combinations 
that correspond to $\sigma\neq \ell$ 
while keeping the 
remaining linear combinations as basis functions 
that correspond to $\sigma=\ell$ and
a well-defined magnetic quantum number $m_\ell$ leads to the so-called 
spherical GTO basis sets. Since the algorithms for 
evaluating the Hamiltonian matrix elements between 
basis functions are based on Cartesian GTOs, the 
use of spherical GTOs involving the corresponding 
transformations is computationally costly.

In most cases, an isotropic particle-particle interaction 
potential commutes with all symmetry elements of the one-particle 
Hamiltonian. Then,
the standard group-theoretical methods can be applied 
straightforwardly to obtain symmetry-adapted configurations 
from symmetry-adapted one-particle orbitals, see, e.\,g., 
Ref.~\cite{cold:schn13}.
The demands become particularly high in situations where atomic or molecular 
wave functions are delocalized across multiple sites in a tweezer array. 
These challenges are further amplified if the array consists of wells that 
are arbitrarily distributed in space --- and potentially vary in 
shape --- i.\,e., if the wells lack the symmetric arrangement 
characteristic of a periodic optical lattice.
Since the exploratory example considered here is 
spherically symmetric, in contrast to tweezer arrays that 
are the real goal of the developed approach, no
symmetry considerations are implemented
in the present calculation.
Still, the obtained results automatically exhibit the proper symmetries. Therefore, as
concluded from the results below, both the one-particle wavefunctions 
$\phi$ as well as the two-electron states $\Psi$ possess 
a well-defined one-particle angular momentum $\ell$ 
or total two-particle angular momentum $L$, respectively.

The Bosonic or Fermionic character, or possibly (sets of) 
distinguishable particles, as well as a possible spin multiplicity,
need
to be accounted for 
the system under 
consideration. While this may still create  
some challenges 
in the case of many particles, the expansion in (anti-)symmetrised 
products of one-particle functions 
allows, nevertheless,
for a 
more straightforward implementation than the use of, e.\,g., 
basis functions expressed in Jacobian coordinates.
Remarkably,
for the case of spin-1/2 Fermions many efficient approaches 
have been developed in the context of electronic-structure 
codes.
%
%
%
%
%
%
\subsection{Two-particle integrals} \label{subsec:integrals}
%
%
%
%
%
\subsubsection{Required integrals}
The calculation of the eigenenergies and the  
eigenfunctions in Eq.~\eqref{eq:system:schrodinger} 
that describe ultracold atoms in a tweezer array 
adopting the CI ansatz \eqref{eq:method:AWF} 
requires the construction of the corresponding 
Hamiltonian matrix in the 
basis of the configurations $\Phi$. Since these 
configurations are expressed as products of the 
single-particle functions $\phi$, see \Eqref{eq:method:WF}, 
the matrix elements of the Hamiltonian (and the 
overlap) between these single-particle functions 
$\phi$ are required.

First, there are the 
three-dimensional one-particle integrals 
$\langle\phi_{\tau}|\op{O}|\phi_{\chi}\rangle$,
where $\op{O}$ is the identity operator (overlap), the 
operator of the kinetic energy, or the one of the 
external potential. The 
kinetic-energy integrals are identical to those occurring 
in any electronic-structure calculation that adopts GTOs. 
If the potential created by the tweezer array is expressed 
either as a polynomial or as a superposition of Gaussians 
with negative prefactors, the one-particle potential-energy 
integrals are basically identical to an overlap integral, but 
between three instead of two GTOs. Again, these integrals 
can be very efficiently evaluated, since the product of 
two Cartesian Gaussians renders as a new  
single Cartesian Gaussian.

Second, there are the 
six-dimensional two-particle integrals 
\begin{eqnarray}\label{eq:method:interaction-integral}
	I &=& 
    \langle\phi_{\tau}\phi_{\sigma} 
    \vert  
    \op{U} 
    \vert  \phi_{\chi}\phi_{\epsilon}\rangle \nonumber \\
    &=&\! \! \! \iint \hspace{-0.1cm} {\rm d}\mathbf r_1\, 
    {\rm d}\mathbf r_2\, \phi_\tau(\mathbf r_1)\phi_\rho(\mathbf
    r_2)
    \,U({r}_{\rm 12}) \,
    \phi_\chi(\mathbf r_1)\phi_\epsilon(\mathbf r_2) 
\end{eqnarray}
that can involve up to four different GTOs characterised by 
the parameters 
$(i,k,m,\tau,\mathbf A)$,
$(j,l,n,\chi,\mathbf B)$,
$(i^\prime,k^\prime,m^\prime,\rho,\mathbf C)$,
and 
$(j^\prime,l^\prime,n^\prime,\epsilon,\mathbf D)$, 
see \Eqref{eq:method:GTOs}. 
These two-particle integrals are challenging, since 
the GTOs are formulated in the absolute Cartesian 
coordinates $\mathbf r_1$ and $\mathbf r_2$ of the 
two particles, but the isotropic interaction potential 
depends on the relative coordinate 
$r_{12} = |\mathbf r_1 -\mathbf r_2|$. In contrast to the 
electronic-structure problem, the interaction potential 
$U$ is not a pure Coulomb potential, but should be 
a short-range
atom-atom interaction potential in the case of ultracold 
atoms. Most importantly, it should once more be reminded 
that already
within
a mean-field calculation a tremendous 
amount of two-particle integrals needs
to be computed. 
This is even worse in the case of a CI calculation. 
Consequently, an efficient and sufficiently accurate 
evaluation of the two-particle interaction integrals 
is of crucial importance. 

In this work, the Morse interaction potential is 
employed. It has a long history in the parametrization 
of the atom-atom interactions, i.\,e., of Born-Oppenheimer 
or spectroscopically obtained potential curves of 
diatomic molecules \cite{gen:herz50,gen:mors29}. 
There are different ways to write the Morse potential 
with corresponding differences in the definition of its 
parameters. Here, the Morse potential is used in the form 
     \begin{align}\label{eq:potential:MRP}
    	{U}(r_{\rm 12})=D_{\rm e}\left( \e^{-2a_{\rm m}( r_{\rm 12}-R_{\rm
        m})} -2\,\e^{-a_{\rm m}( r_{\rm 12}-R_{\rm m})} \right) ,
     \end{align}
where $D_{\rm e}$ is the dissociation energy, 
$R_{\rm m}$ defines the location of the potential minimum, 
and $a_{\rm m}$ defines the width of the potential. 
Despite of having only these three fit 
parameters, the Morse potential usually provides very 
good fits for almost all electronic bound states of 
diatomic molecules, especially for 
their electronic ground states. Furthermore, if inserted 
into the time-independent Schr\"odinger equation describing 
the nuclear motion of a diatomic molecule within the Born-Oppenheimer 
approximation, it possesses known analytical solutions 
representing the rotational and vibrational degrees of 
freedom. In fact, the method employed here for the 
case of a Morse interaction potential
can be rather straightforwardly extended to other atom-atom 
interaction model potentials like the Lennard-Jones 
potential~\cite{gen:buck38} or even numerically provided 
(for example Born-Oppenheimer) potentials, as
will be described in a forthcoming work.

%
%
%
%
\subsubsection{Reduction to one-dimensional form}\label{subsubsec:reduction}
%
%
%

A general formalism for solving integrals
of the type given in \Eqref{eq:method:interaction-integral} was introduced in
the Refs.~\cite{math:silk15a,gen:balc17}.
Thus, here, 
only a brief summary is provided to present
the reader
the general idea and 
notation.

Applying the McMurchie-Davidson scheme \cite{math:mcmu78} 
to the product of four Cartesian GTOs in 
\Eqref{eq:method:interaction-integral} 
potentially 
located at four different centers
yields a product of two 
Cartesian Gaussians.
These Gaussians, 
which are related to particles 1 and 2, respectively,
have the exponents 
$p= \tau + \chi$ and $q= \rho + \epsilon$,
and are centered 
at the positions
$\mathbf P = \frac{\tau \mathbf A + \chi \mathbf B}{p}$ and 
$\mathbf Q = \frac{\rho \mathbf C + \epsilon \mathbf D}{ q}$.
Since the integration has to be performed over the complete 
space, it is, without loss of generality, possible to shift 
the origin of the coordinate system to agree to the 
center of one of the two Gaussians, leading to $\mathbf P=0$.
Then, with the help of the
Gauss-Hermite
functions,
\Eqref{eq:method:interaction-integral}
can be expressed as \cite{gen:balc17}
\footnote{
Note, there are typos in the Eqs.~(11) and (12)
of Ref.~\cite{gen:balc17}. In both equations 
the term $E^{mn}_{u}$ should be
$E^{mn}_{v}$. Likewise, in Eq. (12) the term 
$E^{m^\prime n^\prime}_{u^\prime}$ should be $E^{m^\prime 
n^\prime}_{v^\prime}$.}
     \begin{align}\nonumber\label{eq:method:int I via E}
     	I=&
     	\sum_{t=0}^{i+j}E^{ij}_{t}
     	\sum_{u=0}^{k+l}E^{kl}_{u}
     	\sum_{v=0}^{m+n}E^{mn}_{v}
     	(-1)^{t+u+v}\\ \nonumber
     	&\times \sum_{t^\prime=0}^{i^\prime+j^\prime}E^{i^\prime j^\prime}_{t^\prime}
     	\sum_{u^\prime=0}^{k^\prime+l^\prime}E^{k^\prime l^\prime}_{u^\prime}
     	\sum_{v^\prime=0}^{m^\prime+n^\prime}E^{m^\prime n^\prime}_{v^\prime}\\ 
     	&\times R^{t+t^\prime, u+u^\prime, v+v^\prime} ,
     \end{align}
where the expansion coefficients, for instance $E^{ij}_{t}$,
can be efficiently computed with the help of
recursion relations
satisfied by the GTOs~\cite{math:sams02}.
The most demanding term in 
Eq.~\eqref{eq:method:int I via E}
is the $R$ tensor. Following a special 
case of the Hobson theorem~\cite{gen:hobs65}, and introducing $Q=|\mathbf{Q}|$,
the $R$ tensor can be expanded
as \cite{gen:balc17}
    \begin{align}\nonumber\label{eq:appendix:finalRtensor}
         R^{tuv}=&\sum_{l=0}^{l_{\rm max}}\sum_{m=-l}^{l} \, \textit{c}_i(l,m,t,u,v) \, 
         Z_{lm}
         (\hat{Q}) \sum_{k=0}^{k_{\rm max}}
         \, d_{k}^{l,k_{\rm max}}\\
         &\times Q^{l_{\rm max}-l-2k} \left( D_Q^{l_{\rm max}-k} B \right) ,
     \end{align}
where~$l_{\rm max}=t+u+v$ and 
$k_{\rm max}=\frac{1}{2}(l_{\rm max}-l)$. 
$\hat{Q}$ is a unit vector in the direction of $\mathbf{Q}$ 
and $D_Q=Q^{-1}\partial_Q$. The so-called inverse Schlegel
coefficients $c_i(l,m,t,u,v)$ are given in, e.\,g.,
the Ref.~\cite{math:schl95}, and the coefficients 
$d_k^{l,k_{\rm max}}$ are given in 
the Ref.~\cite{gen:balc17}.
Hence, for practical purposes, both kinds of coefficients
can be pre-computed and stored as a look-up table.
Furthermore, the terms $Z_{lm}(\hat{Q})$ are the real solid spherical harmonics,
and the term $B$ is a function called the basic integral
given by
\begin{align}\nonumber\label{eq:appendix:firstBasicintegral}
      B =& \sqrt{\frac{\pi^5}{p+q}}\, \frac{1}{pq}\int_{0}^{\infty}{\rm d} r_{\rm 12} \,
      {U}({r}_{\rm 12}) \,  r_{\rm 12} \\
      &\times \left[ \frac{\e^{-\frac{pq}{p+q}( r_{\rm 12}-Q)^2}\, -\, \e^{-\frac{pq}{p+q}
      ( r_{\rm 12}+Q)^2}}{Q} \right].
\end{align}

Most computational time is 
devoted to evaluate
the derivatives of the basic integral, $D_Q^{\nu}B$,
with $\nu=l_{\rm max}-k$, since they are
needed in Eq.~\eqref{eq:appendix:finalRtensor}.
For moderate and large values of $Q$, 
an effective recursive relation exists and is given in
the Ref.~\cite{math:silk15a}. 
For small $Q$, this recursion becomes numerically unstable,
and, following the suggestion in Ref.~\cite{math:silk15a},
the $R$-tensor and the basic integral $B$
are evaluated by expanding $B$ in a Taylor series decoupling the derivatives 
from the integration in \Eqref{eq:appendix:firstBasicintegral}. 
%
%
%
%
%
%
%
%
\subsubsection{Special case $Q=0$} \label{subsec:Q=0}
Besides the 
$Q\neq 0$ cases discussed so far, 
$Q=0$ also occurs
if $\rho\mathbf{C}=-\epsilon \mathbf{D}$ (or if $\tau\mathbf{A}=-\chi \mathbf{B}$,
and the coordinate
system is chosen such that 
$\mathbf{Q}=0$). The condition $Q=0$ is especially met, 
if all four GTOs are centered at the same place, 
i.\,e.\,, for $\mathbf{A}=\mathbf{B}=\mathbf{C}=\mathbf{D}=0$. 
As in the case of electronic-structure calculations, the  
accurate description of atoms in a tweezer array requires 
to adopt a number of GTOs centered at every potential 
well. Therefore, a large number of two-particle integrals 
with $Q=0$ needs to be solved. It is thus important to 
consider this case explicitly. In fact, for $Q=0$ the 
algorithm sketched in
Section~\ref{subsubsec:reduction} for $Q\ne 0$ cannot 
be used in a straightforward way as $Q$ appears in 
derivatives and denominators in \Eqref{eq:appendix:finalRtensor}.
Alternatively,
one way to evaluate the integrals in this 
case is to adopt the relations for $Q\neq 0$ and to extrapolate 
them to zero,
which is rather straightforward. However, in view of the expected large
number of integrals that fulfill $Q=0$, a more resource-saving evaluation of
these integrals has been developed and implemented together with the 
$Q\neq 0$ cases discussed above. In fact, in the course of the present work 
it turned out that the special implementation of the $Q=0$ integrals is 
also important, since it increases the numerical accuracy. 

To obtain a solution for $Q=0$, we start with the Taylor series 
of $B$ obtained in Ref.~\cite{math:silk15a} for small values of $Q$  
\footnote{Here, a factor of 2 was
missing in Eq.~(45) of Ref.~\cite{math:silk15a},
which was subsequently
corrected in Eq.~(25) of Ref.~\cite{gen:balc17}.}
\begin{align}\nonumber\label{eq:appendix:finalBasicintegral}
      B =& 4\pi \left( \frac{\pi}{p+q}\right)^{3/2} \e^{-\xi Q^2} \sum_{n=0}^{\infty}
      \frac{(2\xi Q)^{2n}}{(2n+1)!} \\
      &\times \int_{0}^{\infty}{\rm d} r_{\rm 12} \, {U}({r}_{\rm 12}) \, 
       r_{\rm 12}^{2n+2} \, \e^{-\xi  r_{\rm 12}^2} \, ,
\end{align}
where
$\xi=\frac{pq}{(p+q)}$.
To solve the expression for the $R$ tensor
given by
\Eqref{eq:appendix:finalRtensor},
the derivative over the above integral $B$ has to be calculated. 
Since the derivative is with respect to $Q$,
it needs to be applied on the $Q$ dependent terms in 
$B$, i.\,e., $ D_Q^{\nu}B \propto D_Q^{\nu}\,(Q^{2n}\, \e^{-\xi Q^2})$. 
Following repeated recursion, the $\nu^{th}$ derivative can be expanded as
     \begin{align}\nonumber\label{eq:appendix:finaldifferential}
     D_Q^{\nu}(Q^{2n}\e^{-\xi Q^2})=& \sum_{s=0}^{\min(\nu,n)} \binom{\nu}{s} 
     (-2\xi)^{\nu-s}
     \frac{(2n)!!}{(2\nu-2s)!!}\\
     & \times Q^{2\nu-2s}\e^{-\xi Q^2}.
     \end{align}
Thus,
the terms other 
than the inverse Schlegel coefficients 
and the $d_k^{l, k_{\rm max}}$ coefficients
in~\Eqref{eq:appendix:finalRtensor}
can be simplified, along with the 
derivative of~\Eqref{eq:appendix:finaldifferential}, as  

     \begin{align}\nonumber\label{eq:appendix:sphericalfinaldifferential}
     \lim_{Q \to 0}\left[ Z_{lm}(\hat{Q})\, Q^{l_{\rm max}-l-2k} \, D_Q^{l_{\rm max}-k}
     (Q^{2n}\e^{-\xi Q^2}) \right] = \qquad \\
     \sqrt{\frac{2l+1}{4\pi}}  \binom{\frac{l_{\rm max}}{2}}{n}  
     (-2\xi)^{\frac{l_{\rm max}}{2}-n}   (2n)!!  \delta_{\rm m0}  
     \delta_{\rm l0}  \delta_{\rm k\frac{l_{\rm max}}{2}}\,,
     \end{align}
where the Kronecker deltas~$\delta_{ij}$ arise because
only the coefficients of $Q^0$ survive.
Hence, plugging 
\Eqref{eq:appendix:sphericalfinaldifferential} 
and the remaining terms of $B$ in \Eqref{eq:appendix:finalBasicintegral}
into \Eqref{eq:appendix:finalRtensor},
and applying the Kronecker deltas, the $R$ tensor can be 
written as
     \begin{align}\nonumber\label{eq:appendix:Rtensor}
     	R^{tuv} =& \frac{2\pi^2}{(p+q)^{\frac{3}{2}}} \,
     	\textit{c}_i(0,0,t,u,v) \,
     	d^{0,\frac{l_{\mx}}{2}}_{\frac{l_{\mx}}{2}} \, 
     	\sum_{n=0}^{\frac{l_\mx}{2}}
     	\biggl\{(-1)^{\frac{l_\mx}{2}-n}\\
        & \times \binom{\frac{l_\mx}{2}}{n} \,
     	\frac{(2\xi)^{n+\frac{l_\mx}{2}}}{(2n+1)!!} \, \Pi_{2n+2}(\xi)
     	\biggr\}\ ,
     \end{align}
which vanishes for odd values of $l_\mx = t+u+v$.
As previously mentioned,
the inverse Schlegel coefficients $c_i$ and the 
$d_{l_{\rm max}/2}^{0, l_{\rm max}/2}$
coefficients can be straightforwardly computed.
As a consequence,
the only demanding term present in \Eqref{eq:appendix:Rtensor}
is $\Pi_{2n+2}(\xi)$, which may be defined as
     \begin{align}\label{eq:appendix:Baseintegral}
     	\Pi_{\lambda}(\xi)=\int_0^\infty {\rm d} r_{\rm 12}\,  r_{\rm 12}
        ^{\lambda} \, {U}( r_{\rm 12}) \, \e^{-\xi  r_{\rm 12}^2} 
     \end{align}
with $\lambda=2n+2$.
A further simplification of
\Eqref{eq:appendix:Baseintegral} strongly depends on the 
shape of the interaction potential $U( r_{\rm 12})$. 
Since the equation is one dimensional, 
we refer to it as the \textit{master integral}.
Considering the explicit example of the Morse interaction 
potential in \Eqref{eq:potential:MRP} and inserting it into
\Eqref{eq:appendix:Baseintegral}, 
the master integral can be written as
\begin{equation}\label{eq:appendix:Morseintegralfinal}
     	\Pi_{\lambda}(\xi) \!  = \! 
        D_{\rm e} \e^{-a_{\rm m}R_{\rm m}} \!  \! \left[
        \e^{-a_{\rm m}R_{\rm m}} S(\lambda, \! - 2 a_{\rm m},\xi)
        \!  - \!  2 S(\lambda, \! - a_{\rm m},\xi) \right] \;. 
\end{equation}
Here, $S$ is an integral that is described in the next section.
%
%
%
%
%
%
%
%
\subsubsection{$S$ Integral} \label{subsec:S-Integral}
The $S$ integral in its generic form is defined as 
     \begin{align}\label{eq:appendix:Sintegral}
     	S(\alpha,\beta, \gamma) = \int_0^\infty {\rm d}x \ x^\alpha \, \e^{\beta x - 
        \gamma x^2} ,
     \end{align}
where $\alpha$, $\beta$, and 
$\gamma$ are real numbers with $\gamma > 0$.
Assuming $\alpha$ to be a non-negative integer, three 
different results are obtained depending on $\beta$,     
\begin{widetext}
\begin{eqnarray}
S(\alpha,\beta,\gamma) &=& 
\left\{
\begin{array}{l}
\frac{\Gamma\left(\alpha+1\right)}{(2\, \sqrt{\gamma})^{\alpha+1}} \,
     		\mathcal{U}\left(\frac{\alpha+1}{2},\frac{1}{2},\frac{\beta^2}{4\gamma}\right)
            \textnormal{, if $\beta<0$,}
            \label{eq:appendix:U} \\
\frac{\Gamma\left(\frac{\alpha+1}{2}\right)}{2\, 
            \gamma^{\frac{\alpha+1}{2}}}
            \textnormal{, if $\beta=0$,} 
            \label{eq:appendix:Gamma}\\
\frac{1}{2\,\gamma^{\frac{\alpha}{2}+1}}  
            \bigg[ \beta \,\Gamma\left(\frac{\alpha}{2}+1\right)          
     		\mathcal{M}\left(\frac{\alpha}{2}+1,\frac{3}{2},\frac{\beta^2}{4\gamma}\right)  
            +
            \sqrt{\gamma}\, \Gamma\left(\frac{\alpha +1}{2}\right) 
     		\mathcal{M}\left(\frac{\alpha+1}{2},\frac{1}{2},\frac{\beta^2}
            {4\gamma}\right)\bigg]  
            \textnormal{, if $\beta>0$} \,. 
            \label{eq:appendix:M}
\end{array}
\right.
\end{eqnarray}
\end{widetext}
Here, $\Gamma$ is the Gamma function, while $\mathcal{M}$
and $\mathcal{U}$ are the Kummer's and Tricomi's
confluent hypergeometric functions,
respectively~\cite{math:silk15a}. 
Note that in Ref.~\cite{math:silk15a} 
only the relation for $\beta < 0$ 
is explicitly given and thus it is implied that it 
remains valid for all values of $\beta$. This is the case for $\beta=0$, 
though inserting $\beta=0$ yields the compact result given 
on the second line of \Eqref{eq:appendix:Gamma}. If all GTOs are 
centered at the same 
location and the Morse potential is adopted to describe the 
interatomic interaction, as in the case in this work,
only~$\beta<0$ occurs. 
However, if the GTOs are 
centered at different locations,
for the Morse potential as well as 
in the case of other particle-particle 
interaction potentials like the Lennard-Jones or the Gaussian 
potentials (both have already been implemented, but will be 
reported elsewhere) also $\beta>0$ occurs.
All in all, the solution for $\beta>0$ was
derived and is already implemented for 
the planned future applications to tweezer arrays.
%
%
%
%
%
%
%
\section{Computational details} \label{sec:details}
As mentioned above,
a system of two interacting ultracold Bosonic atoms trapped in a harmonic potential is 
considered, since accurate reference data can be obtained
for validation of the results using the approach described in 
Refs.~\cite{cold:gris09,cold:gris11,cold:schn13}. 
The length scale of the trap is defined in
harmonic-oscillator units as 
$d_{\rm ho}=\sqrt{\hbar/\mu \omega}$.
The energies are given in units of $\hbar\omega$ with the trap 
frequency~$\omega$. Despite the fact that in the present 
approach absolute coordinates are adopted, the reduced mass 
$\mu=\frac{m_{1}m_{2}}{m_{1}+m_{2}}$
is used in the discussion, because the reference approach 
adopts the coordinates of relative motion and the 
center-of-mass motion.
In fact, in many theoretical studies of trapped 
two-atom systems, the coupling of center-of-mass motion and 
relative motion is ignored and only the solutions of the 
relative motion are discussed, since within this approximation 
the center-of-mass motion is trivial.
With no loss of generality, $\hbar=m=\omega=1$ is chosen, 
since the present study is only meant as a proof of principles. 
This choice of parameters leads to $\hbar \, \omega = 1$.

\begin{figure}[h!]
    	\centering
        \includegraphics[scale=0.5]{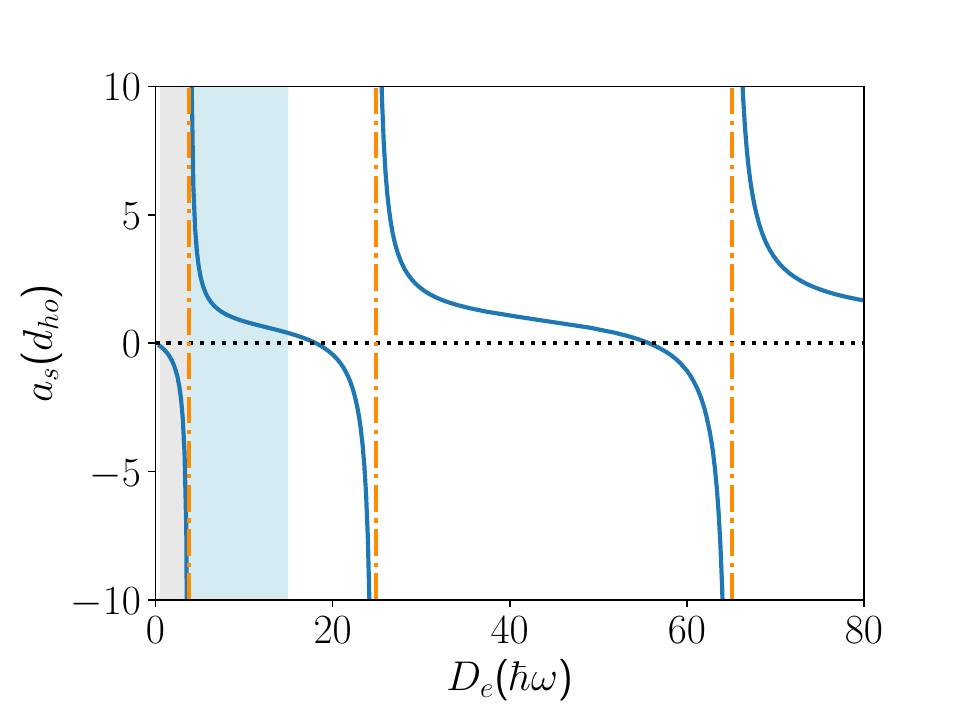}
    	\captionof{figure}{Scattering length $a_{\rm s}$ (continuous blue line) 
        of the Morse potential
        given by Eq.~\eqref{eq:potential:MRP} for
        $R_{\rm m}=0.212 \,d_{\rm ho}$ and 
        $a_{\rm m}=\sqrt{20} \,d^{-1}_{\rm ho}$
        as a function of the
        potential depth $D_{\rm e}$.
        The poles (vertical dashed-dotted orange)
        mark the positions
        where an additional bound state emerges as $D_{\rm e}$ increases.
        The gray and light blue shaded areas mark the $D_{\rm e}$ range considered
        in this work, where the first bound state appears in the light blue region.
        }
     \label{figure:results:scatteringlength}
    \end{figure}

Experimentally, the interactions between ultracold atoms can be almost  
arbitrarily tuned with the help of magnetic Feshbach resonances
\cite{cold:chin10,aies:ayma78,cold:loft02,cold:gris10}. 
Furthermore, due to the low temperatures
the effective interactions among Bosons are dominated
by $s$-wave scattering and can be characterized by a
single parameter, i.\,e.,
the $s$-wave scattering length ($a_{\rm s}$) \cite{cold:wein99}.
In this work, the $s$-wave scattering length is tuned by a  
variation of the depth of the Morse  
potential $D_{\rm e}$ while keeping
$R_{\rm m}=0.212 \, d_{\rm ho}$ 
and $a_{\rm m}=\sqrt{20}\, d^{-1}_{\rm ho}$
constant.
The scattering-length as a function of the variation of the 
Morse-potential depth is shown in
\figref{figure:results:scatteringlength}.
The regions of negative and positive scattering lengths correspond 
to an effectively attractive or repulsive interaction between the 
two atoms, respectively.
Note, at the poles (orange dash-dotted lines)
occurring for $D_{\rm e}=[3.823,~24.877,~65.075]\,\hbar\omega$,
the scattering length diverges, changing
from infinitely strong attraction ($a_{\rm s}\rightarrow -\infty$) to 
infinitely strong repulsion 
($a_{\rm s}\rightarrow +\infty$). 
These poles mark the positions of the Feshbach
resonances where an additional molecular bound state appears, since
the number of molecular bound states that are 
supported increases with the depth of the particle-interaction 
potential. In the calculations performed in this work, the
range $D_{\rm e} \in [0.5-15.0\,]\,\hbar\omega$
(gray and light blue shaded areas) is considered.
Within the gray region,
the Morse potential is too shallow 
to support any bound state.  
However, the first bound state appears 
in the area marked by the light blue color.

In this exploratory study,
the employed basis set comprise of $\sigma=i+k+m=0,1,2$, and $3$.
Specifically, 4 GTOs for each corresponding orbital type
(e.\,g. s, p$_x$, p$_y$, p$_z$, etc.) are adopted.
This yields 4 GTOs for $\sigma=0$, 12 GTOs for $\sigma=1$, 24 GTOs for 
$\sigma=2$, and 40 GTOs for $\sigma=3$, adding up to a total 
number of $N_{\rm GTO}=80$ functions. 
This basis set is henceforth named GTO basis. For simplicity, 
the same 4 values of the exponents $\tau$ in \Eqref{eq:method:GTOs} 
are adopted, independently of $\sigma$
(see discussion in the Sec.~\ref{subsec:convergence} for details). 
With $N_{\rm GTO}=80$ and no use of symmetry, 
$N_{\rm CF}=3240$ configurations $\Phi$ are obtained which
is, thus,
the dimension of the CI matrix. Since the configurations 
are directly expanded in (non-orthogonal) GTOs, a generalized 
eigenvalue problem needs to be solved. 
For validation purposes, two different LAPACK diagonalization
routines, DSYGV and DSYEV, are employed. 
The results obtained from both routines are found to be in perfect agreement 
for the leading 12 digits.

Our calculations show a high sensitivity to numerical errors in the
evaluation of the basic integral~$B$
[see \Eqref{eq:appendix:firstBasicintegral}]. 
Any numerical error propagates and accumulates through the computation 
of the sums in Eqs.~\eqref{eq:method:int I via E}
and~\eqref{eq:appendix:finalRtensor}.
The $S$ integral can be calculated with the aid of the hypergeometric functions, 
cf.~\Eqref{eq:appendix:M}.
Hence, a fast and accurate evaluation of these functions was mandatory. 
Introducing a representation for the inverse Schlegel coefficients 
$c_i(0,0,t,u,v)$ and the 
$d^{0,l_{\mx}/2}_{l_{\mx}/2}$ coefficients for the particular case 
$Q=0$ reduced the number of sums and function calls in the $R$ tensor, 
cf. Eqs.~\eqref{eq:appendix:Rtensor} and \eqref{eq:appendix:finalRtensor}.  
This reduced the computational time and improved the accuracy.

In the reference calculations, we used 52 $B$ splines for
both relative
and center-of-mass computations,
with maximum 
values of angular 
momentum $\ell_{\rm max}=6$ and the magnetic number $|m_{\ell,{\rm max}}|=6$ 
for the spherical harmonics
(see Refs.~\cite{cold:gris11, cold:sala16a, cold:cape23}
for further details on
the reference approach).
%
%
%
%
%
%
%
%
%
%
\section{Numerical results}\label{sec:results}
This section presents the numerical results of our study.
First,
Sec.~\ref{subsec:convergence}
provides a convergence study.
Second,
Sec.~\ref{subsec:spectrum}
compares the energy spectrum obtained using the GTOs basis set 
with reference results adopting a well-established method
based on $B$ splines and spherical harmonics. 
Third, Sec.~\ref{subsec:eigenfunctions}
provides a comparative study of eigenfunctions produced by both 
approaches. 
%
%
%
%
%
%
\subsection{Convergence study} \label{subsec:convergence}
For a first validation of the present approach and its implementation,  
the convergence of the ground-state energy as a function of 
the number of included basis functions is investigated.
Four different particle-particle-interaction strengths are 
considered, i.\,e., the Morse-potential depths 
$D_{\rm e}= [3,~5,~10,~13]\, \hbar\omega$ that 
correspond to the scattering-length values   
$a_{\rm s}=[-2.550,\, 2.758,\, 0.852, \,0.579]\, d_{\rm ho}$
(cf.~\figref{figure:results:scatteringlength}), respectively.
Note that for $D_{\rm e}= 3.0 \,\hbar\omega$ no bound state is supported, while  
the remaining values chosen for $D_{\rm e}$ support a single bound 
state.

As previously mentioned,
an efficient structure calculation requires a careful selection of the 
basis-set parameters. In the case of Cartesian GTOs, 
these parameters are the position vectors ${\bf A}$ of the Gaussians,
the chosen types of GTOs, i.\,e., the values $i,k,$ and $m$, and 
the damping exponents $\tau$ [see \Eqref{eq:method:GTOs}].
The basis must be able to adequately cover the one-particle 
Hilbert space. Since the system considered here
comprises only a single-well trap potential, GTOs centered at 
the origin are adopted ($\bf A=0$ for all GTOs). 
Linear combinations of GTOs with 
appropriately chosen basis parameters are the eigenfunctions of 
the harmonic trap. Therefore, the description of non-interacting 
particles in a harmonic trap would be straightforward with the 
present approach. However, the proper description of the interaction  
between the atoms, especially if they form molecular bound states, is 
challenging when adopting trap-centered GTOs. 
Thus, various combinations 
of four different exponents $\tau$ were tested with respect to 
their impact on the ground-state energy.
Small values of the exponents $\tau$ correspond to diffuse 
basis functions, while increasing values of $\tau$ lead to 
more localized ones.   

  	 \begin{figure}[h!]
 	   \centering
 	   \includegraphics[scale=0.5]{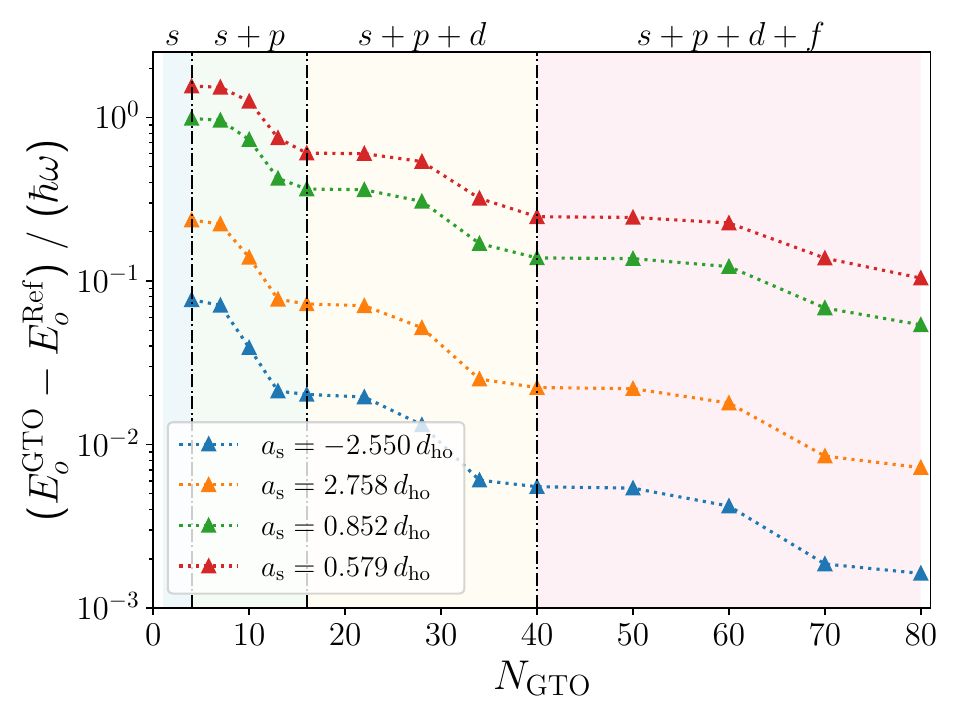}
 	   \caption{
       The difference between the molecular 
       ground state energy $E^{\rm GTO}_{\rm 0}$
       computed with Gaussian-type Orbitals (GTOs)
       and the corresponding 
       reference value $E^{\rm Ref}_{\rm 0}$ 
       as a function of the number 
       of GTOs, $N_{\rm GTO}$, for the scattering-length 
       values given in the legend.
       The colored shaded areas (and the labels on the top axis)
       indicate the types of GTOs
       (s for $\sigma=0$, 
       p for $\sigma=1$, 
       d for $\sigma=2$,  and
       f for $\sigma=3$)
       that are included.}
     \label{fig:results:convergence}
      \end{figure}

Since even a particle-particle interaction that is isotropic 
with respect to the relative-motion coordinate usually breaks 
the single-particle symmetry, the single-particle solutions 
of different irreducible representations
are coupled to an irreducible representation of the total system.
For the case study,
the $\ell$ and $m_{\ell}$ single-particle quantum numbers
couple to the total angular momentum $L$ (and $M_L$) of the 
two-atom system.
Therefore, the addition of configurations containing orbitals 
with different angular momentum $\ell$ is needed for converged CI 
results, as they describe the so-called angular correlation.  
This applies, e.\,g.,
also to the ground state in which all 
particles occupy --- even if the particle-particle interaction is 
included within the mean-field approximation 
--- the lowest orbital which in the present isotropic case has 
$\ell=0$ symmetry. As a consequence,
the convergence 
behavior with respect to the exponents $\tau$ and the included 
types of GTOs ($\sigma$ values) is not independent of each other 
and, in principle, the exponents $\tau$ need to be optimized 
whenever new GTOs are added to the calculation.
Noticeably, the exponents $\tau$ need to be optimized separately also for
each inter-particle interaction and for every state, if accurate results 
are required. In contrast, in the present exploratory study the exponents 
$\tau$ were optimized with the goal of finding one set of values that 
provides overall the lowest ground-state energies for the four scattering-length
values considered. Based on a simple parameter variation, the best compromise for
the GTO exponents were found to be $\tau= 0.3, 0.5, 1.0\,$ and 2.2.

The convergence behavior of the numerically obtained ground-state energies 
$E_0^{\rm GTO}$ as a function of the number $N_{\rm GTO}$ of 
included Cartesian GTOs is shown in Fig.~\ref{fig:results:convergence} 
for the four different scattering-length values $a_{\rm s}$. The energies 
are given relative to the quasi-exact reference values 
$E_0^{\rm Ref}$, i.\,e., $E_0^{\rm GTO} - E_0^{\rm Ref}$.    
Starting with a minimal basis set formed only 
with four s-type GTOs,
the number of GTOs, $N_{\rm GTO}$, 
is increased by adding successively four p-, d-, and f-type Cartesian GTOs, 
each of them comprising three, six, and ten GTOs, respectively.
Consistent with the variational principle, all energies $E_0^{\rm GTO}$ 
lie above the corresponding reference values $E_0^{\rm Ref}$ and the  
energy differences decrease monotonically with an increasing value of $N_{\rm GTO}$. 

Remarkably, the energy differences as a function of $N_{\rm GTO}$ 
of all the curves exhibit similar shapes, 
though some of them differ by almost two orders of magnitude.
Best convergence is found for $a_{\rm s}=-2.550 \,d_{\rm ho}$. 
This is understandable, since the corresponding Morse potential with 
depth $D_{\rm e}=3.0\,\hbar\omega$ is 
not deep enough to support a molecular bound state. Therefore, the 
ground-state wave function is a slightly distorted
trap state due to the non-zero (attractive) interatomic interaction. The second 
best convergence is achieved for the potential depth  
$D_{\rm e}=5.0\,\hbar\omega$. In this case, the ground state corresponds to a 
loosely bound molecular bound state. Therefore, 
the deviation from a trap state is larger than for 
$D_{\rm e}=3.0\,\hbar\omega$, but still comparatively small. As a consequence, the 
wavefunction is expected to be still very similar to the trap 
ground state. Thus, this state is well described by a relatively small 
number of GTOs. This changes if the particle-particle interaction 
increases, but still only a single molecular bound state is supported. 
With increasingly attractive interaction, the bound state becomes 
more deeply bound and, correspondingly, the wavefunction becomes more 
compact in the relative-motion coordinate, deviating even more from 
the one of a trap state (see Sec.~\ref{subsec:eigenfunctions} for 
details on the wavefunctions). 

The results shown in Fig.~\ref{fig:results:convergence}
confirm the expectation that the energy convergence of the GTO basis set
in absolute coordinates reduces
with the depth of the interatomic interaction potential.
Note, the goal of the present approach is neither the description 
of deeply bound molecular states nor of energetically very 
high-lying trap states. The former states can be efficiently treated 
with (standard) molecular codes, since the trap potentials are 
negligible. The treatment of the latter ones is also straightforward, 
since the interatomic interaction is negligible and thus the problem 
reduces, to a good approximation, to a single-particle problem. 

It is interesting to note that the found convergence 
behavior as a function of $N_{\rm GTO}$ differs nevertheless almost exclusively by 
an overall factor (constant shift on the logarithmic 
scale in Fig.~\ref{fig:results:convergence}), as stated before. 
In fact, the optimization of the set of four $\tau$ values that provides 
the best overall results 
for all values of $D_e$ was clearly dominated by those yielding the 
most deeply bound molecular ground state. As a result, 
the optimization of the GTO exponents $\tau$ was effectively carried out by minimizing 
the ground-state energy for the cases with $a_{\rm s}=0.852\,d_{\rm ho}$ and 
$0.579\,d_{\rm ho}$, while the results for the two other $a_{\rm s}$ values 
remained (automatically) well converged. 

More relevant than the absolute energies are the binding energies, 
here defined as the difference between the corresponding 
non-interacting trap-state energy and the computed ground-state energy,  
{\it i.\,e.}, $E_{\rm binding}= E_{\rm trap} -E_{\rm computed}$ 
with $E_{\rm trap} = 3 \,\hbar\omega $.
The relative errors (in percent) in the binding energies 
obtained for 80 $N_{\rm GTO}$ and the four scattering-length values  
$a_{\rm s}=[-2.550,\, 2.758,\, 0.852, \,0.579]\, d_{\rm ho}$ 
are $[0.315\,\%,\, 0.699\,\%,\, 1.811\,\%,\, 2.316\,\%]$, respectively.

The CPU time for one CI calculation with $N_{\rm GTO}=80$ on 
a single thread of an Intel Core i9-12900K processor is 212 
seconds, 
and remains
basically identical for the four 
interatomic interaction potentials considered in 
\figref{fig:results:convergence}. Thus the CPU
time depends basically only on the chosen basis set. The 
results show that the integrals, especially the ones describing the 
interparticle interactions adopting a realistic Morse potential, can be 
calculated with sufficient efficiency to allow for the theoretical 
treatment of few-atom systems in optical tweezer arrays, as 
evidently much larger basis sets are feasible than the ones used in 
this exploratory study. It may be noted that the number of terms 
in the master integral in \Eqref{eq:appendix:Baseintegral},
two in the case of the here considered Morse potential 
(see \Eqref{eq:appendix:Morseintegralfinal}), 
strongly affects the computational time.
%
%
%
%
%
%
%
%
\subsection{Energy spectrum} \label{subsec:spectrum}
\begin{figure}[h!]
      \centering
      \includegraphics[scale=0.5]{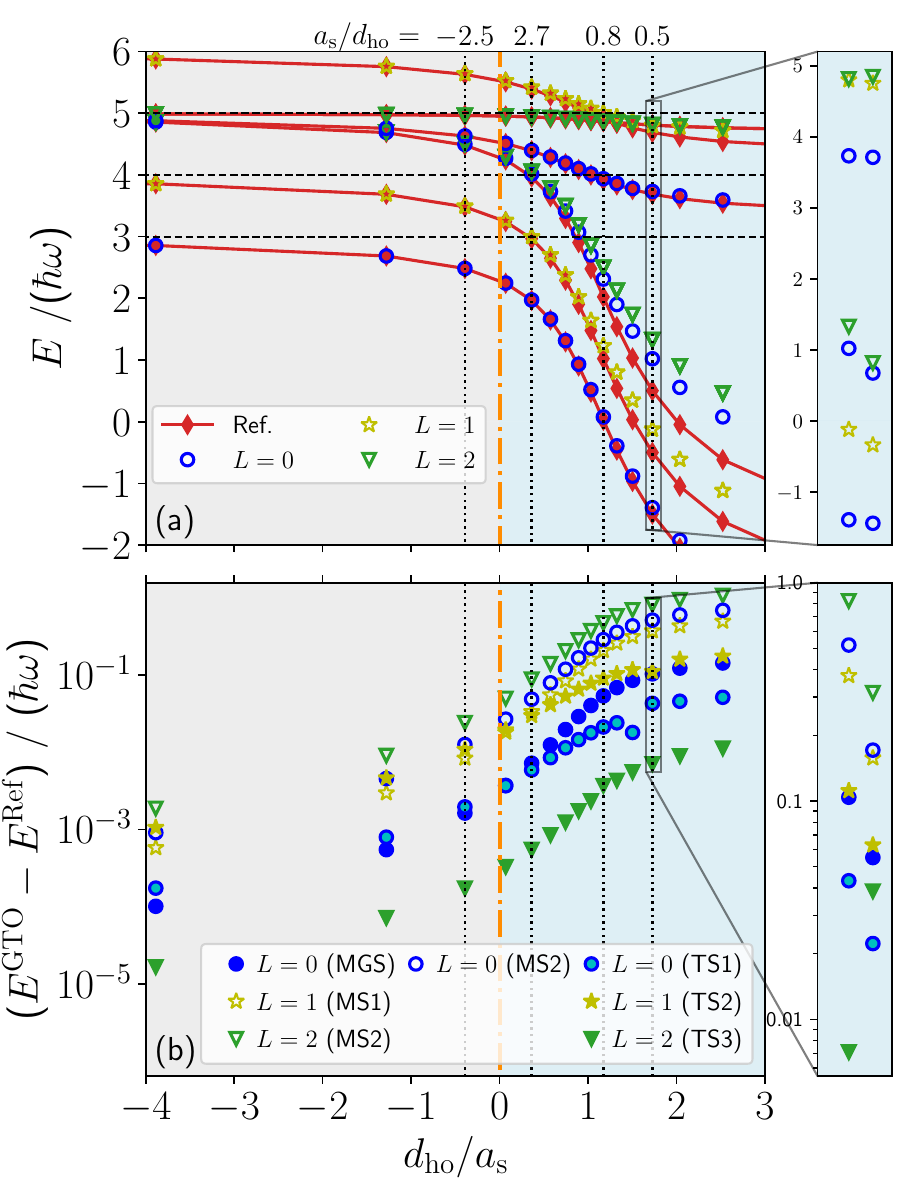}
      \caption{(a) Energies $E$ of some selected states of two atoms 
       in a harmonic trap with total angular momentum $L$ are shown as a function
      of the ratio between the characteristic trap size ($d_{\rm ho}$)
      and the scattering length ($a_{\rm s}$). The results for the GTO basis set, 
      see Sec.\ref{sec:details},  
      are compared to quasi-exact reference results $E_{\rm Ref}$. 
      The gray and light-blue shaded regions indicate the same 
      $a_{\rm s}$ range as in Fig.~\ref{figure:results:scatteringlength}, 
      while the vertical orange dashed-dotted 
      line marks the resonance position. The dashed horizontal
      black lines show the asymptotic values of trap states 
      in the absence of interatomic interaction. 
      The vertical dotted black lines with the values of
      $a_{\rm s}/d_{\rm ho}$ on the top axis correspond to the  
      four values of $a_{\rm s}$ discussed in the convergence 
      study in \figref{fig:results:convergence}.
      (b) Deviation of the energies $E^{\rm GTO}$ obtained with the 
      present GTO basis from the quasi-exact reference values 
      $E^{\rm Ref}$. Here, MGS stands for the molecular ground 
      state, MS1 and MS2 stand for the $1^{\rm st}$ and the $2^{\rm nd}$
      excited  molecular bound states, respectively,
      while TS1, TS2, and TS3 stand for the trap states.
      The insets on the right provide a comparison 
      of the energies (a) and energy deviations (b) at 
      $a_{\rm s} = 0.579\,d_{\rm ho}$ obtained by the two 
      GTO basis sets GTO ($N_{\rm GTO}=80$, left column) and 
      GTO-2 ($N_{\rm GTO}=96$, right column), 
      shifted horizontally with respect to each other for better 
      readability).}
      \label{fig:results:MRP}
    \end{figure}

The energies of the ground state and some lower-lying excited 
states are shown in \figref{fig:results:MRP} as a function of 
the effective particle-particle interaction. As is common 
practice, this interaction strength is expressed in terms of the 
corresponding \emph{inverse} scattering length $a_{\rm s}$ that 
in turn is given in units of the trap length $d_{\rm ho}$. 
Recall that small 
positive (negative) values of  $d_{\rm ho}/a_{\rm s}$
indicate strong repulsive (attractive) interparticle 
interaction. As already mentioned, in the present 
calculations the trap length is fixed while the scattering length 
is varied by tuning the depth of the Morse potential
(see Fig.~\ref{figure:results:scatteringlength}). 

With the present choice of parameters, the negative values of 
$d_{\rm ho}/a_{\rm s}$ correspond to the case where the 
Morse potential does not support
any bound state.
In the limit of vanishing attractive interaction 
($d_{\rm ho}/a_{\rm s}\rightarrow -\infty$) the ground-state 
energy approaches the one of the lowest lying trap state, 
i.\,e., the ground state of two non-interacting particles 
in a harmonic-oscillator potential (3$\hbar \omega$). 
With increasing depth of the Morse potential describing the 
interparticle interaction ($d_{\rm ho}/a_{\rm s}$ approaching 
zero from below) 
the ground-state energy decreases. As is discussed in 
Sec.~\ref{sec:details} a further increase of the depth $D_e$ 
of the Morse potential leads to a Feshbach resonance due to 
the appearance of a bound state. The scattering length changes 
from $-\infty$ to $+\infty$ at the resonance and thus its 
inverse is continuous around $d_{\rm ho}/a_{\rm s} = 0$, 
as is the ground state energy. The latter decreases further 
with increasing depth of the Morse potential, since the bound 
state becomes more strongly bound. The effective interaction 
between two unbound, colliding atoms in this region is however 
repulsive, the repulsion strength decreases with increasing 
depth of the Morse potential. As the bound state becomes more 
strongly bound, the influence of the trap potential becomes 
less important and finally negligible.

Since the relative motion and the center-of-mass motion 
separate for a harmonic trap, there is an infinite number of 
states separated by $\hbar \omega$ that correspond to the 
relative-motion ground state, but
possess excitations in the center-of-mass motion 
(see Refs.~\cite{cold:gris11, cold:sala16a, cold:cape23}
for further details). 
In \figref{fig:results:MRP}\,(a) the lowest lying 
four of those states are shown. These, in the case of 
positive values of $a_{\rm s}$ bound molecular states are denoted 
as MGS (molecular ground state, no center-of-mass 
excitation, total angular quantum number $L=0$), MS1 
(one quantum of center-of-mass 
excitation, $L=1$), and MS2 (two quanta of center-of-mass 
excitation, $L=0$ and $L=2$). Due to the isotropic harmonic 
trap potential, the MS2 $L=0$ and $L=2$ states are degenerate, 
but as a result of the finite basis set, 
deviations from this degeneracy can be seen, especially 
for large positive $a_{\rm s}$ values. In contrast to 
the exact results (see the results of the reference 
calculation),
the curves of the states MGS, MS1, and MS2 
obtained with the GTO basis differ only approximately by 
a quantum of center-of-mass energy $\hbar\omega$. For 
$a_{\rm s}<0$ the agreement to the reference results 
is very good and all states converge asymptotically 
for $a_{\rm s} \rightarrow -\infty$ to the corresponding 
trap states describing non-interacting particles (their 
energies being indicated by black dashed lines). However, 
there is a visible increasing deviation 
from the reference results for increasing positive values 
of $a_{\rm s}$, see the deviations from the reference 
results shown in \figref{fig:results:MRP}\,(b). 
Furthermore, the 
best agreement is achieved
for the ground state 
MGS. It is similar for the states MS1 $L=1$ and MS2 $L=0$, 
though slightly better for the MS1 state.
The largest deviations  are observed
for the MS2 $L=2$ state. However, the shape 
of the deviation curves is very similar for all  
four states.

Since for the 
considered depths of the Morse potential 
at most one molecular bound state is supported, the states 
with excitation in the relative-motion degree of freedom 
are all trap states and would be unbound continuum states 
in the absence of a trap. The lowest lying trap state 
(here denoted by TS1) 
has total angular momentum $L=0$ and is in the center-of-mass 
ground state of the harmonic potential, but has
two quanta of excitation in the relative motion ($2\,\hbar\omega$). 
Therefore, it lies 
in the limit $a_{\rm s}\rightarrow -\infty$ by about 
the energy difference $\Delta E=2\,\hbar\omega$ 
above the ground state at $E=5\,\hbar\omega$. As the molecular 
bound state emerges for positive $a_{\rm s}$ and the effective 
interaction between the atoms decreases, the TS1 state approaches 
asymptotically for $a_{\rm s}\rightarrow +\infty$ the energy of 
the ground state of two non-interacting atoms in a harmonic 
potential, i.\,e., $E=3\,\hbar\omega$. A very similar behavior 
is seen in \figref{fig:results:MRP} for the state named TS2, 
but the energy curve is shifted by $2\,\hbar\omega$ compared 
to the one of TS1, since the two states differ by 
two quanta of relative-motion excitation. Thus the energies 
differ exactly by this value for all scattering lengths $a_{\rm s}$, 
if a fully converged result is considered.  
In \figref{fig:results:MRP} the state 
TS3 is also shown which possesses the total angular-momentum quantum number $L=2$ 
and is thus five-fold degenerate. 
Its energy shows an even smaller dependence 
on the interatomic interaction than is found for the states TS1 
and TS2.
In fact, TS3 is nearly independent of the 
interatomic interaction. However, a slight deviation is observed for 
increasing positive values of $d_{\rm ho}/a_{\rm s}$. This is due to the fact that
in the ultracold regime, the
$s$-wave scattering is typically dominant, as the centrifugal barrier suppresses 
higher-order partial waves. If the interaction potential is broadened,
the centrifugal barrier is effectively lowered, allowing contributions from 
channels beyond the $s$-wave channel \cite{sala2016}. 
Consequently, the weak deviation of the 
TS3 state from the horizontal line at $E=5\,\hbar\omega$ provides direct evidence of 
the $d$-wave interactions.

At first glance, the deviations of the GTO results from 
the reference ones show a similar qualitative behavior for 
the trap states and the molecular bound states. The accuracy 
decreases as the inverse scattering length varies from 
$-\infty$ to $+\infty$. A slight deviation from the trend is 
noticed for $L = 0$ (TS1) at $d_{\rm ho}/a_{\rm s} = 1.502$ 
and $L = 1$ (TS2) at $d_{\rm ho}/a_{\rm s} = 1.727$. 
This is due to the fact that there are crossings to higher 
bound state(s) at these points, lowering the energy of these 
two trap states (see~\footnote{At these points 
the GTOs energies are lowered due to an avoided 
crossing with higher bound state(s) of the same symmetry, 
which are not shown in the graph. This lower energy 
leads to a better agreement with the quasi-exact results.}). 
However, compared to the molecular bound states, the decrease 
of accuracy is less pronounced for the trap states, especially 
for positive values of the scattering length. Furthermore, the 
TS3 state shows the smallest deviation from the correct result. 
This is in accordance with the earlier mentioned expectation 
that the GTO basis functions are more suitable for describing 
the trap states, while more efforts (basis functions) are 
required to properly describe the atom-atom interaction. 
Therefore, state TS3 with an energy that is almost independent 
on the interatomic interaction can be most accurately described.

In order to obtain more accurate results, the basis-set parameters 
should be optimized separately both for each value of $a_{\rm s}$ and 
for each state. While an extensive study of basis-set convergence is 
beyond the scope of this work, it is nevertheless of interest to 
explore whether and how the basis set could be improved, especially 
in the regime of positive scattering length values where the molecular 
state becomes more deeply bound, {\it i.\,e.}, when the interatomic 
interaction is important. Thus for $a_{\rm s} = 0.579 \,d_{\rm ho}$ 
($d_{\rm ho}/a_{\rm s} = 1.727$) a second basis set was introduced and 
optimized for describing the ground state. Compared to the basis set 
discussed so far (GTO), this second basis set (named GTO-2) contains 
only three instead of four GTOs for $\sigma= 0, 1, 2$, and $3$, as the 
energies had only been marginally improved by the 4th GTO. In this case, 
$\tau = 1.0, 1.7$, and $2.3$ were obtained for the exponents optimized 
for the ground state. Furthermore, basis GTO-2 contains additional GTOs 
with $\sigma = 4$ and $5$ (g- and h-type orbitals) with exponent 
$\tau = 1.0$, resulting in a total of $N_{\rm GTO} = 96$ functions 
and 4656 configurations. 

A comparison of the results obtained with 
GTO and GTO-2 is given in the inset of \figref{fig:results:MRP}. Besides 
state TS3,
all energies obtained with GTO-2 are lower than the ones 
yielded with basis set GTO, see \figref{fig:results:MRP}(a). In 
accordance with the empirical experience that a lower energy corresponds 
to a more accurate result \footnote{The variational principle only applies 
to the lowest state of each symmetry. For excited states it applies 
strictly only, if the state is orthogonal to all lower lying \emph{exact} 
states.}, the errors of all but the TS3 state 
are reduced, if GTO-2 is adopted, see \figref{fig:results:MRP}(b). 
The improvement comes at the cost of substantially increased computational 
time, 929\,s (GTO-2) compared to 212\,s (GTO). However, it proves that 
the energies can be systematically improved, if a correspondingly larger 
basis is adopted. Furthermore, the result highlights the importance 
of angular correlation for describing (strongly) interacting particles 
when adopting a CI approach, since it is the GTOs with larger $\sigma$ 
values that lead to the improvement. While in the present isotropic 
example the inclusion of angular correlation adopting Cartesian GTOs 
with large values of $\sigma$ is rather inefficient, it should be 
emphasized that for the intended applications (non-isotropic tweezer 
arrays) these GTOs are also important for describing the non-isotropic 
shape of the wavefunctions. 

Since the highly
lying TS3 state is only extremely 
weakly affected by the interatomic interaction, it is not fully unexpected 
that more GTOs with small value of $\sigma$ are found to be more important 
for its description than those with large values of $\sigma$. Especially 
the description of higher lying trap states require less compact 
GTOs with diffuse exponents, while the optimization of the ground-state 
energy favors more compact GTOs. On the other hand, especially for very 
high lying states the wavefunctions possess more and more nodes and thus 
also require the inclusion of GTOs with larger $\sigma$ values. Alternatively, 
additional GTOs that are not centered at the minima of the trap potential 
could be used. 

In conclusion, the results indicate that adopting the new approach reasonable 
energy spectra can be obtained with relatively small computational efforts. 
Enlarging the basis set, as well as optimizing the basis-set parameters 
for a given interatomic interaction potential or state, allows for more 
accurate energy spectra.
The findings 
are consistent, thus indicating the 
proper implementation.
%
%
%
%
%
%
%
\subsection{Eigenfunction analysis}\label{subsec:eigenfunctions}
\begin{figure*}[h!]
      \begin{subfigure}[b]{0.30\textwidth}
      \includegraphics[scale=0.30]{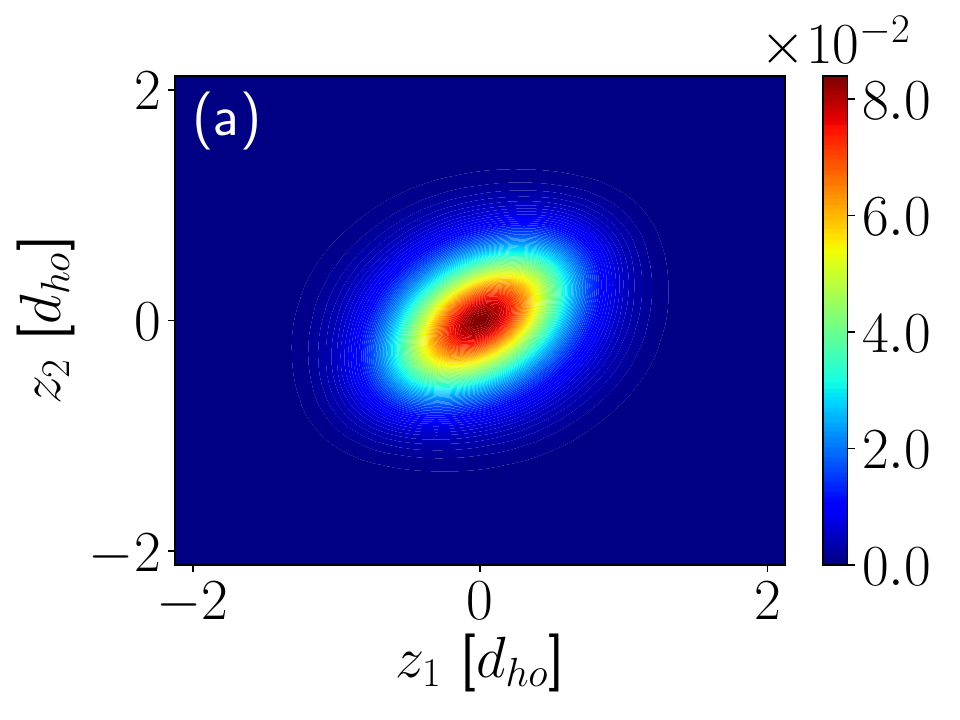}
      \end{subfigure}
      \begin{subfigure}[b]{0.30\textwidth}
	   \includegraphics[scale=0.30]{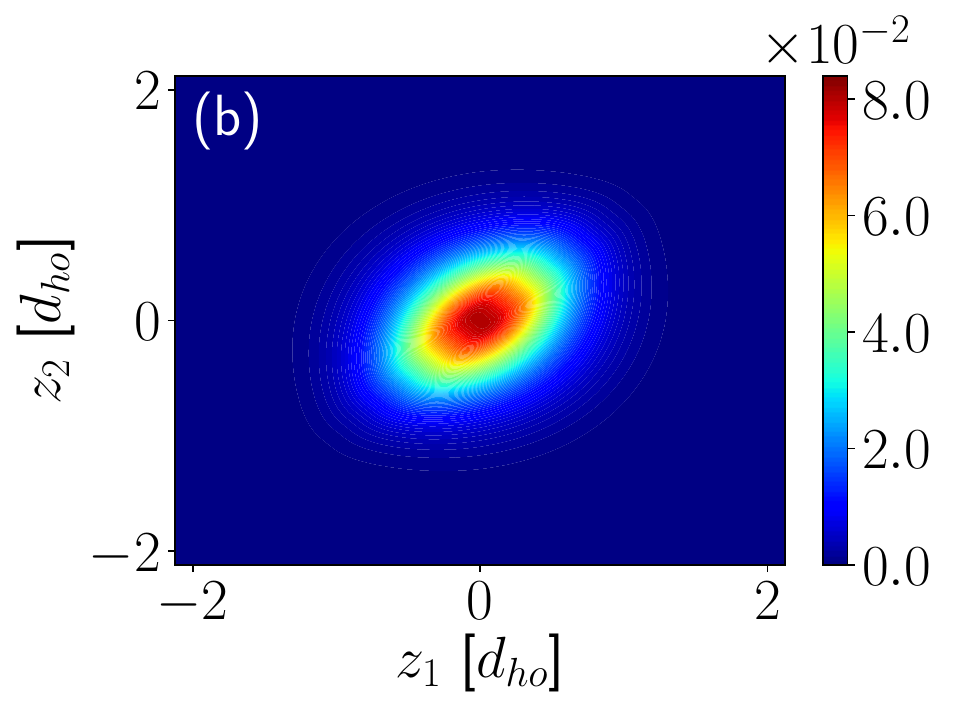}
      \end{subfigure}
      \begin{subfigure}[b]{0.30\textwidth}
	   \includegraphics[scale=0.30]{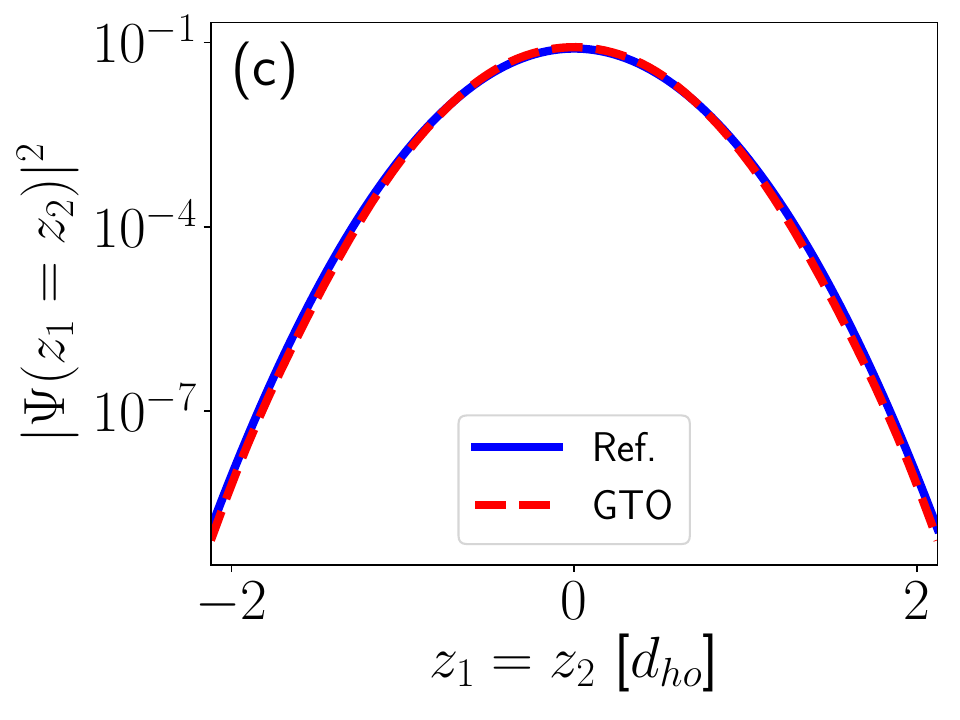}
      \end{subfigure}
      \begin{subfigure}[b]{0.30\textwidth}
      \includegraphics[scale=0.30]{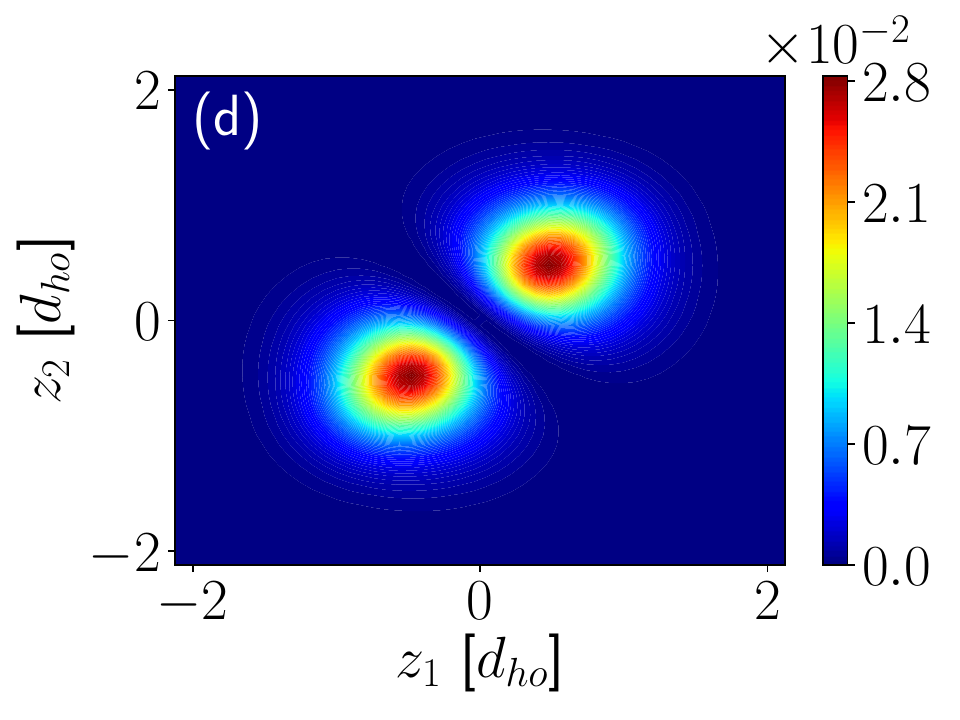}
      \end{subfigure}
      \begin{subfigure}[b]{0.30\textwidth}
      \includegraphics[scale=0.30]{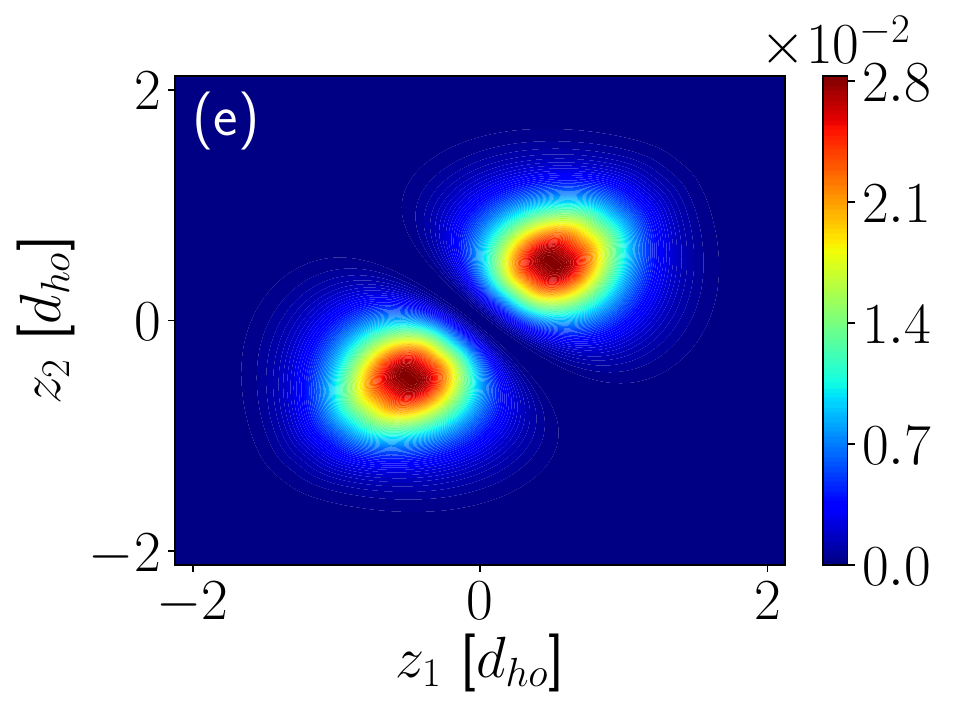}
      \end{subfigure}
      \begin{subfigure}[b]{0.30\textwidth}
      \includegraphics[scale=0.30]{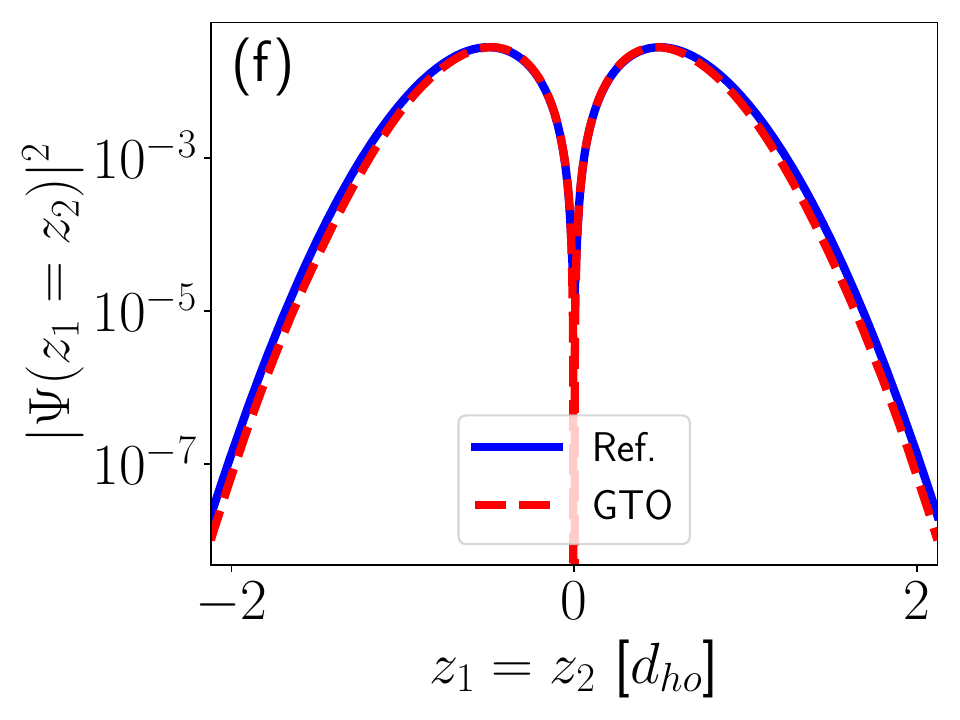}
      \end{subfigure}
      \begin{subfigure}[b]{0.30\textwidth}
      \includegraphics[scale=0.30]{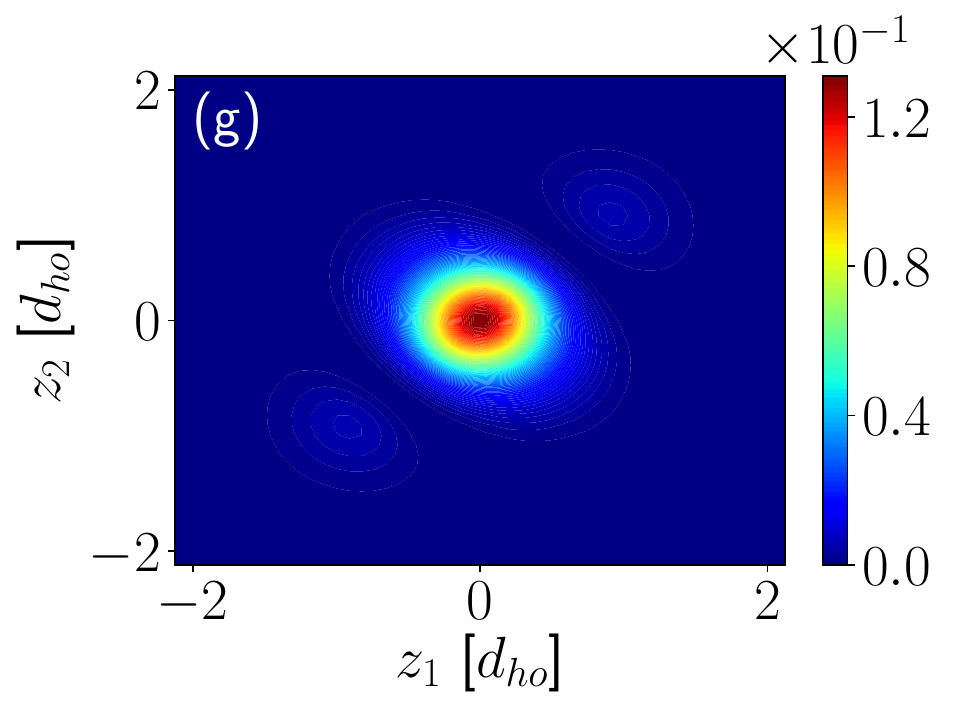}
      \end{subfigure} 
      \begin{subfigure}[b]{0.30\textwidth}
      \includegraphics[scale=0.30]{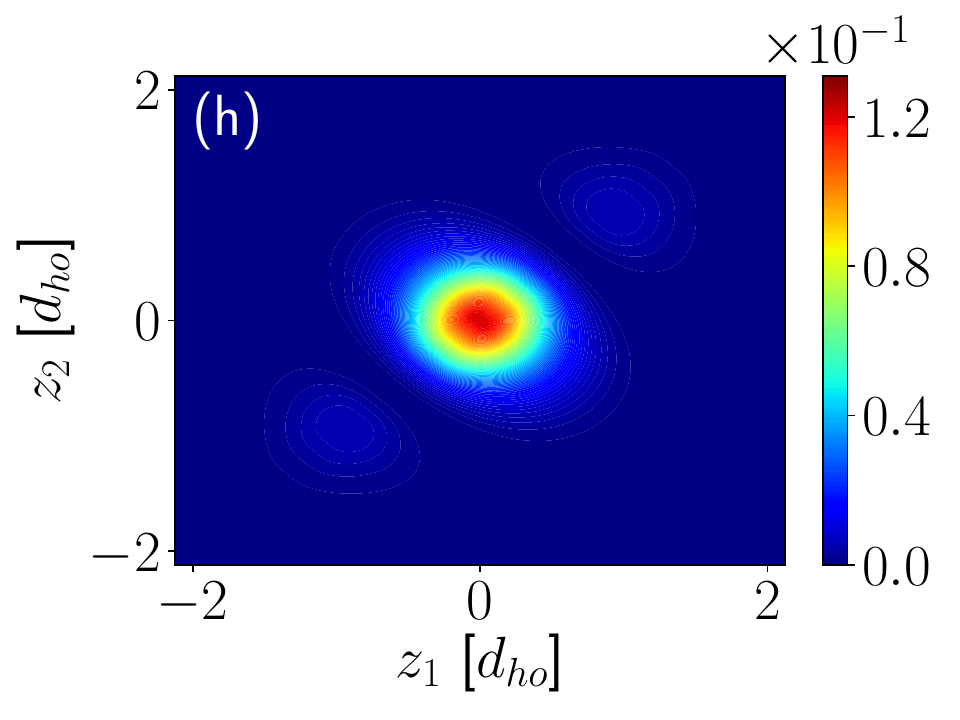}
      \end{subfigure}
       \begin{subfigure}[b]{0.30\textwidth}
      \includegraphics[scale=0.30]{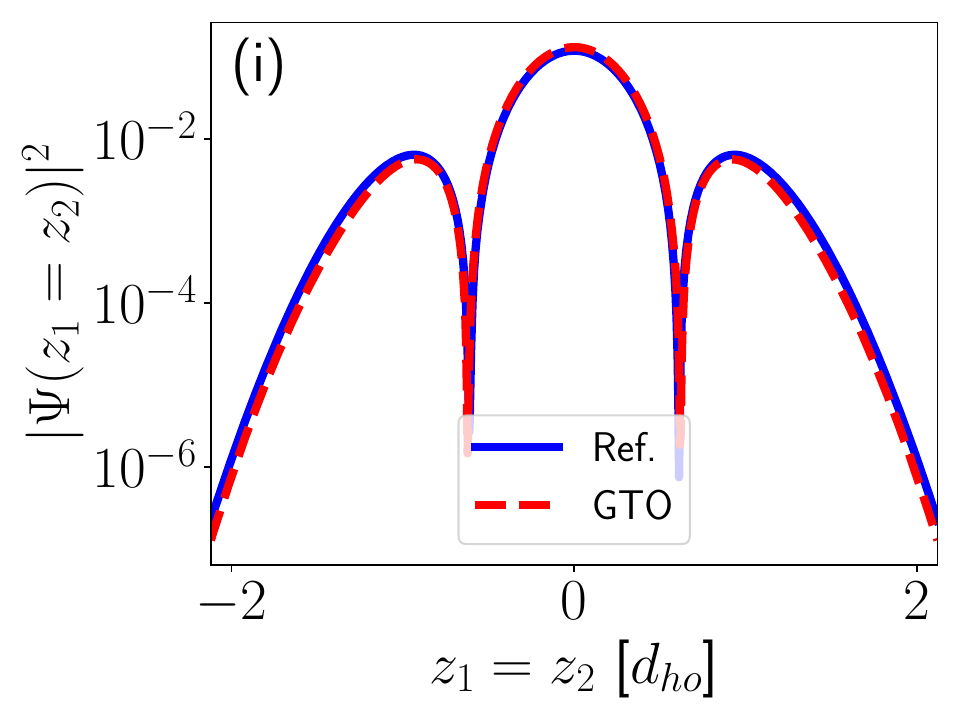}
      \end{subfigure}
	   \caption{Cuts through the six-dimensional 
       probability density (squared wave function)
       along the [$z_1,z_2$] plane
       ($|\Psi(z_1,z_2; x_1=x_2=y_1=y_2=0)|^2$)
       for an attractive interaction characterized by
       $d_{\rm ho}/a_{\rm s} = -0.392$.
       First row (a,b,c): MGS state ($L=0$)
       (a) GTO approach ($E^{\rm GTO}=2.484\,\hbar\omega$),
       (b) reference result ($E^{\rm Ref}=2.483\,\hbar\omega$), 
       and
       (c) one-dimensional cut along the diagonal $z_1=z_2$ for both, GTO 
       approach and reference.
       Second row (d,e,f): as 1st row, but for MS1 state 
       ($L=1$, $E^{\rm GTO}=3.491\,\hbar\omega, E^{\rm Ref}=3.483\,\hbar\omega$). 
       Third row (g,h,i): as 1st row, but for MS2 
       ($L=0$, $E^{\rm GTO}=4.495\,\hbar\omega, E^{\rm Ref}=4.483\,\hbar\omega$).
       }   
       \label{fig:results:wave functions}
      \end{figure*}

In this section, the six-dimensional
eigenfunctions obtained with the present GTO approach are 
discussed and compared with those produced with the reference 
approach in relative-motion coordinates 
based on $B$ splines and spherical harmonics.  
\figref{fig:results:wave functions} shows cuts along 
the [$z_1,z_2$] plane ($x_1=x_2=y_1=y_2=0$) through the
probability densities of the MGS, MS1, 
and MS2 states for an attractive interaction 
characterized by $d_{\rm ho}/a_{\rm s} = -0.392$,
which corresponds to the potential 
depth~$D_{\rm e}=3.0\,\hbar\omega$.

Figures~\ref{fig:results:wave functions}(a) and
\ref{fig:results:wave functions}(b) depict the 
cuts through the MGS probability (total angular momentum 
$L = 0$, see~\figref{fig:results:MRP}), obtained with 
either the GTO approach or the reference calculation, 
respectively. For a more quantitative comparison,
\figref{fig:results:wave functions}(c)
shows in addition the one-dimensional cut through the density along 
the diagonal $z_1=z_2$. For symmetry reasons, the ground state MGS 
is nodeless. Both atoms preferentially stay together in 
the middle of the trap (maximum at $z_1=z_2=0$). 
Very good agreement is found for the results obtained either with 
the GTO approach or the reference calculation. 

Figures~\ref{fig:results:wave functions}(d) to (f) show 
the corresponding results for the MS1 state ($L=1$). The 
two-dimensional cut through the probability density 
possesses two lobes for $z_1=z_2\neq 0$) as well as a 
nodal plane along the antidiagonal ($z_1=-z_2$). Consequently, 
there is a node in the one-dimensional cut at $z_1=z_2=0$. 
Due to their attractive interaction, the two atoms still 
stay preferentially together, but not in the middle of the 
trap. The agreement with the reference result is again very 
good, though slightly worse than for the ground state, 
as is visible in  \figref{fig:results:wave functions}(f) 
especially away from the maxima, {\it i.\,e.}, for larger 
absolute values of $z_1=z_2$.
A similar result is obtained for the state MS2 ($L=0$),
whose probability density shows two nodal planes 
in the two-dimensional cut presented in \figref{fig:results:wave functions}(h),
and two nodes in the one-dimensional cut
in \figref{fig:results:wave functions}(i). 
(Due to the 
finite numerical resolution adopted for plotting, the two 
nodes have a finite value.) While the 
overall agreement with the quasi-exact reference results 
is still very good, there are already small, but visible 
differences at the maxima.
%



\onecolumngrid

\twocolumngrid

\begin{figure*}[h!]
	   \centering
      \begin{subfigure}[b]{0.3\textwidth}
      \centering
	   \includegraphics[scale=0.3]{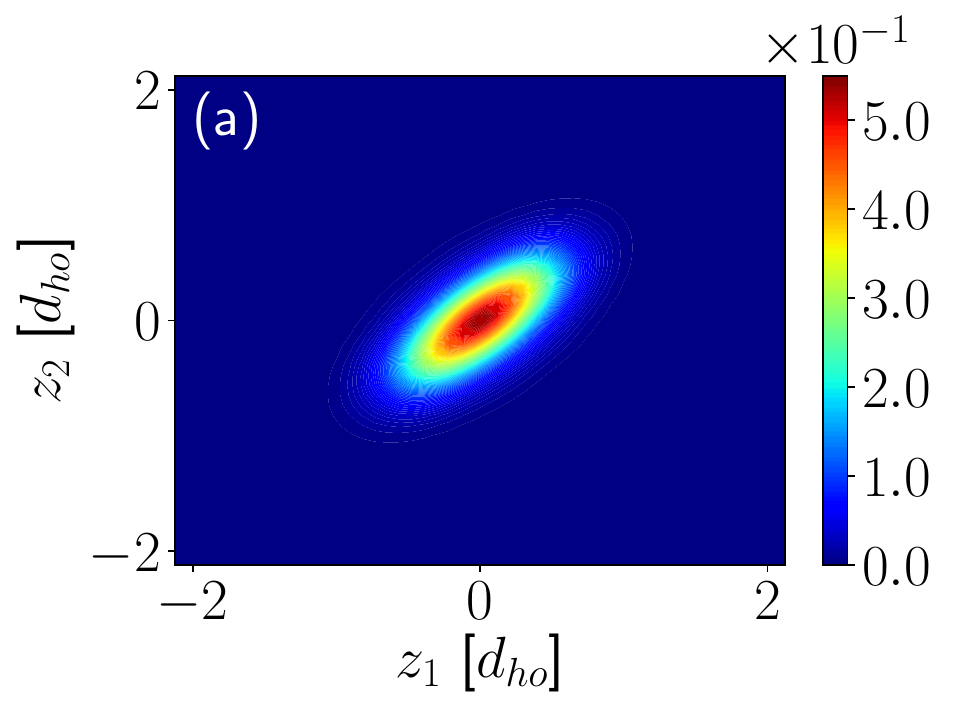}
      \end{subfigure}
      \begin{subfigure}[b]{0.3\textwidth}
      \centering
	   \includegraphics[scale=0.3]{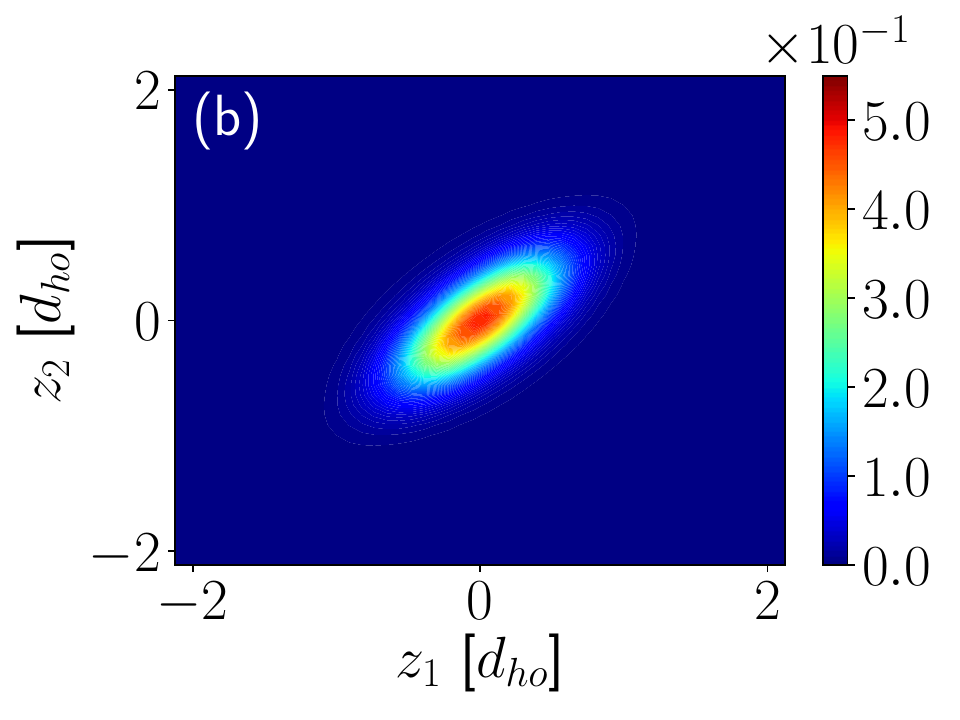}
      \end{subfigure}
      \begin{subfigure}[b]{0.3\textwidth}
      \centering
      \includegraphics[scale=0.3]{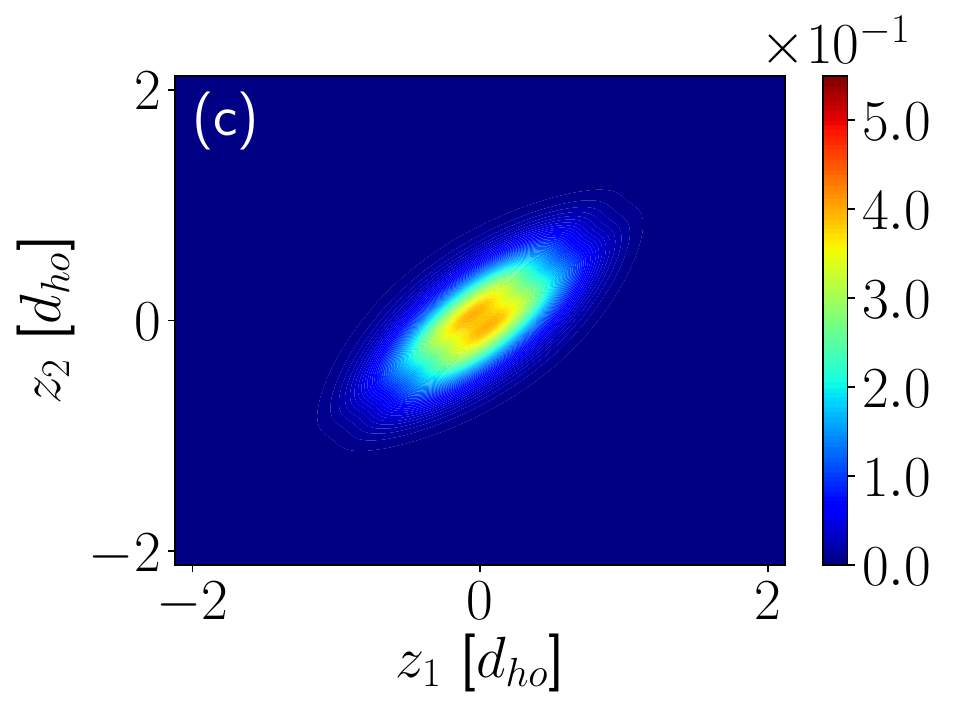}
      \end{subfigure}
      \begin{subfigure}[b]{0.3\textwidth}
      \centering
      \includegraphics[scale=0.3]{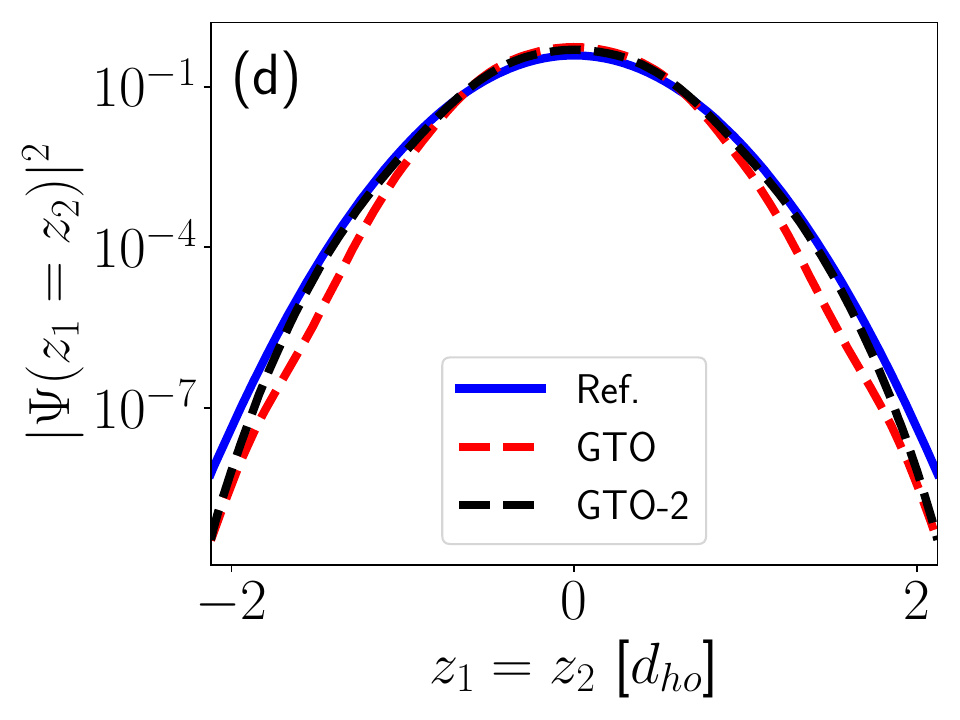}
      \end{subfigure}
      \begin{subfigure}[b]{0.3\textwidth}
      \centering
      \includegraphics[scale=0.3]{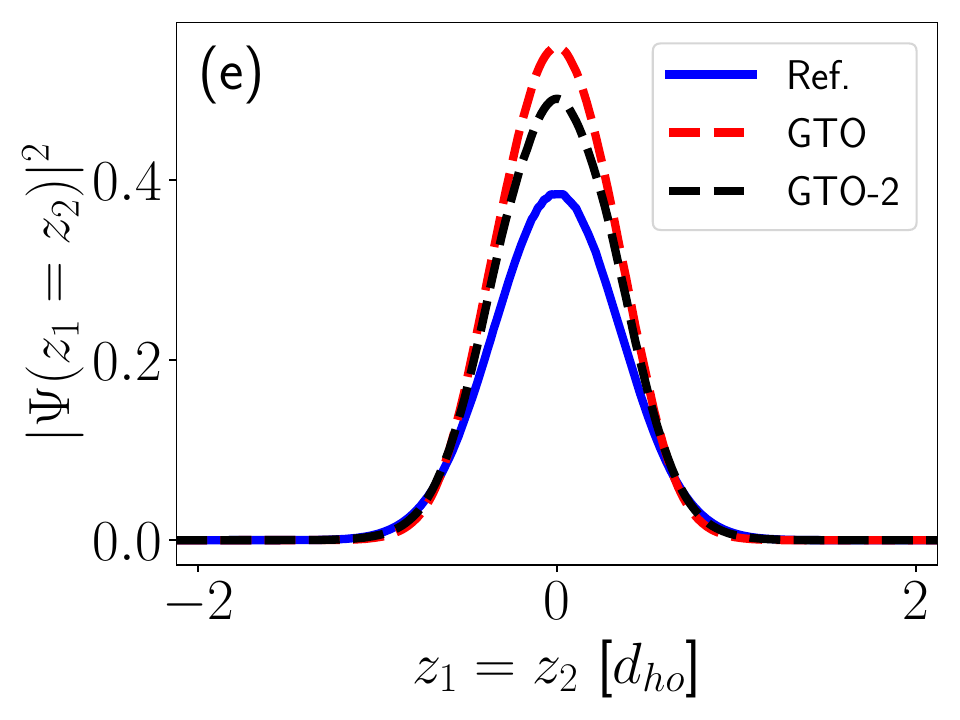}
      \end{subfigure}
      \begin{subfigure}[b]{0.3\textwidth}
      \centering
      \includegraphics[scale=0.3]{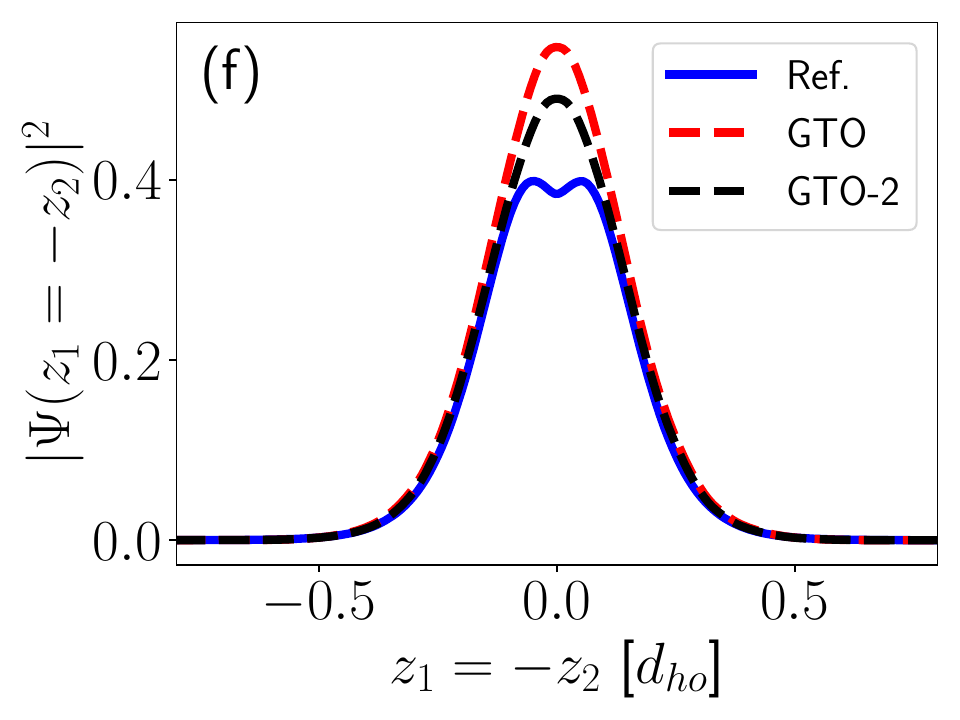}
      \end{subfigure}
	   \caption{
       Same as \figref{fig:results:wave functions}, but for a more 
       deeply-bound ground state MGS ($d_{\rm ho}/a_{\rm s} = 1.727$).
       Upper row: 
       (a) GTO basis set ($E=-1.393\,\hbar\omega$),
       (b) GTO-2 basis set ($E=-1.442\,\hbar\omega$),
       and
       (c) reference result ($E=-1.497\,\hbar\omega$).
       Lower row: one-dimensional cuts through the probability 
       densities, 
       (d) along the diagonal $z_1=z_2$ (log scale),
       (e) along the diagonal $z_1=z_2$ (linear scale), 
       and
       (f) along the anti-diagonal $z_1=-z_2$ (linear scale).
       } 
      \label{fig:results:wave functionsGTOsBsplinesGS}
      \end{figure*}

As has been discussed in \secref{subsec:spectrum}, the accurate 
description of the states becomes more challenging, if (large)  
positive values of $d_{\rm ho}/a_{\rm s}$ and thus repulsive 
interatomic interaction is considered. For example, the ground-state 
energy for $d_{\rm ho}/a_{\rm s} = 1.727$ was shown to be much 
closer to the quasi-exact reference result, if basis set GTO-2 
instead of GTO was adopted, since basis functions with larger 
$\sigma$ values (angular momenta) were required for capturing 
a larger fraction of the angular correlation. In 
\figref{fig:results:wave functionsGTOsBsplinesGS} 
the earlier introduced two- and one-dimensional probability
densities of the ground state (MGS) are shown for basis GTO 
as well as GTO-2, and compared with the reference results.  
Already the two-dimensional cuts ($z_1,z_2;x_1=x_2=y_1=y_2$) 
show a reasonable agreement, but also clearly visible 
differences. Again, a more quantitative picture is obtained 
when looking at the one-dimensional cuts along the the 
diagonal ($z_1=z_2$). While the deviation to the reference 
result is already visible for the logarithmic plot, it becomes 
even more evident on the linear scale.  In accordance with the 
improved energy, also the density obtained with GTO-2 compares 
better with the reference density than the GTO result, but the 
deviation remains rather pronounced. 

A more careful look at the two-dimensional densities reveals 
that the GTO results show a rather smooth function with a 
single maximum. However, the reference density shows some 
substructure in the center. In fact, there are two maxima 
slightly offset from the center. This becomes more transparent 
in the one-dimensional cut along the anti-diagonal ($z_1=-z_2$) 
shown in \figref{fig:results:wave functionsGTOsBsplinesGS} (f). 
While the reference result shows a small dip in the center, this 
dip is absent in both, the GTO as well as the GTO-2 densities. 
Thus even the GTO-2 basis is not capable of reproducing this 
dip that is a consequence of the interatomic interaction.

\onecolumngrid

\twocolumngrid

It needs to be emphasized that the shown densities are pair correlated 
(conditional) densities. Similar to most electronic-structure calculations 
correlation is also here a relatively small effect and the energy of a 
state depends mostly on the one-particle density. This explains why the 
energy, especially for the ground state and $d_{\rm ho}/a_{\rm s} = 1.727$ 
is, especially when using GTO-2, rather well agreeing with the reference 
result, despite the rather large deviations discussed in the context of 
\figref{fig:results:wave functionsGTOsBsplinesGS}. 

%
%
%
%
%
%
%
%
%
%
\section{Conclusion and outlook} \label{sec:conclusions}
In this work, a new numerical method was developed, implemented, and tested  
that heads for the study of trapped few-body systems (indistinguishable Bosons 
and Fermions or distinguishable particles) beyond the mean-field 
approximation by means of configuration interaction (exact diagonalization). 
The approach aims for the description of more than two particles 
in multi-centered trap potentials, especially ultracold atoms in 
optical tweezer arrays. Cartesian Gaussians in absolute coordinates, well 
established in electronic-structure calculations, are used as basis functions. 
The six-dimensional particle-particle integrals were formulated and 
implemented for a realistic diatomic interaction potentials, namely the 
Morse potential. Since the evaluation of the large number of six-dimensional 
particle-particle interaction integrals is the bottleneck of structure 
calculations, some efforts were spent to find an efficient, but nevertheless 
sufficiently accurate implementation of the integral calculation. In order 
to validate the proper implementation of the integrals, their accuracy, and 
the computational speed, a full configuration interaction calculation of 
two particles (Bosons) trapped in an isotropic harmonic trap with variable 
inter-particle interaction strength has been performed, since for this system 
quasi-exact results can be obtained. Furthermore, a first investigation of 
the suitability of Gaussian basis functions has been performed. This included 
a rudimentary convergence study, also in order to cross-check the results, 
as an increased basis set should provide lower ground-state energies according 
to the variational principle.

The obtained results and their comparison to quasi-exact reference data 
appears consistent, indicating a correct formulation and implementation 
of the integrals as well as the two-particle full CI calculation. The 
speed of the integral calculation indicates that larger systems of 
interest are accessible, especially if the integral calculation is 
parallelized, possibly on graphics cards, as is planned for the future.

The integrals presented in this work form the heart of a currently developed code 
that allows for both, mean-field and beyond mean-field calculations describing 
a variable number of particles (identical or non-identical Bosons or Fermions) 
trapped in an multi-centered potential of arbitrary geometry. This will be used 
for studies of the influence of confining potentials like optical tweezer 
arrays on, e.\,g., atomic clocks, qubits made from neutral atoms, or phenomena 
like Efimov states. Furthermore, inelastic confinement resonances can be studied 
for more than two particles.

%
%
%
%
\section*{ACKNOWLEDGMENTS}
        MuR acknowledges the HEC Overseas Scholarships for
        PhD in Selected Fields, Phase III, Batch-1, 2020 (57558560) provided 
        in collaboration with the DAAD. 
        This work has been partially supported by the Grants PID2021-122711NB-C21 funded 
        by MCIN/AEI/10.13039/501100011033 
        and NSF PHY-2309135 granted to the Kavli Institute for Theoretical Physics (KITP). 
        The authors acknowledge helpful discussions with Robert Moszynski and Michal Lesiuk.
   
%
%
%
%
%

%
%
%
%

\begin{thebibliography}{74}
\expandafter\ifx\csname natexlab\endcsname\relax\def\natexlab#1{#1}\fi
\expandafter\ifx\csname bibnamefont\endcsname\relax
  \def\bibnamefont#1{#1}\fi
\expandafter\ifx\csname bibfnamefont\endcsname\relax
  \def\bibfnamefont#1{#1}\fi
\expandafter\ifx\csname citenamefont\endcsname\relax
  \def\citenamefont#1{#1}\fi
\expandafter\ifx\csname url\endcsname\relax
  \def\url#1{\texttt{#1}}\fi
\expandafter\ifx\csname urlprefix\endcsname\relax\def\urlprefix{URL }\fi
\providecommand{\bibinfo}[2]{#2}
\providecommand{\eprint}[2][]{\url{#2}}

\bibitem[{\citenamefont{Spar et~al.}(2022)\citenamefont{Spar, Guardado-Sanchez,
  Chi, Yan, and Bakr}}]{cold:spar22}
\bibinfo{author}{\bibfnamefont{B.~M.} \bibnamefont{Spar}},
  \bibinfo{author}{\bibfnamefont{E.}~\bibnamefont{Guardado-Sanchez}},
  \bibinfo{author}{\bibfnamefont{S.}~\bibnamefont{Chi}},
  \bibinfo{author}{\bibfnamefont{Z.~Z.} \bibnamefont{Yan}}, \bibnamefont{and}
  \bibinfo{author}{\bibfnamefont{W.~S.} \bibnamefont{Bakr}},
  \bibinfo{journal}{Phys. Rev. Lett.} \textbf{\bibinfo{volume}{128}},
  \bibinfo{pages}{223202} (\bibinfo{year}{2022}).

\bibitem[{\citenamefont{Pause et~al.}(2023)\citenamefont{Pause, Preuschoff,
  Sch{\"a}ffner, Schlosser, and Birkl}}]{cold:paus23}
\bibinfo{author}{\bibfnamefont{L.}~\bibnamefont{Pause}},
  \bibinfo{author}{\bibfnamefont{T.}~\bibnamefont{Preuschoff}},
  \bibinfo{author}{\bibfnamefont{D.}~\bibnamefont{Sch{\"a}ffner}},
  \bibinfo{author}{\bibfnamefont{M.}~\bibnamefont{Schlosser}},
  \bibnamefont{and} \bibinfo{author}{\bibfnamefont{G.}~\bibnamefont{Birkl}},
  \bibinfo{journal}{Phys. Rev. Res.} \textbf{\bibinfo{volume}{5}},
  \bibinfo{pages}{L032009} (\bibinfo{year}{2023}).

\bibitem[{\citenamefont{Bloch et~al.}(2023)\citenamefont{Bloch, Hofer, Cohen,
  Browaeys, and Ferrier-Barbut}}]{cold:bloc23}
\bibinfo{author}{\bibfnamefont{D.}~\bibnamefont{Bloch}},
  \bibinfo{author}{\bibfnamefont{B.}~\bibnamefont{Hofer}},
  \bibinfo{author}{\bibfnamefont{S.~R.} \bibnamefont{Cohen}},
  \bibinfo{author}{\bibfnamefont{A.}~\bibnamefont{Browaeys}}, \bibnamefont{and}
  \bibinfo{author}{\bibfnamefont{I.}~\bibnamefont{Ferrier-Barbut}},
  \bibinfo{journal}{Phys. Rev. Lett.} \textbf{\bibinfo{volume}{131}},
  \bibinfo{pages}{203401} (\bibinfo{year}{2023}).

\bibitem[{\citenamefont{Levine et~al.}(2019)\citenamefont{Levine, Keesling,
  Semeghini, Omran, Wang, Ebadi, Bernien, Greiner, Vuleti{\'c}, Pichler
  et~al.}}]{cold:levi19}
\bibinfo{author}{\bibfnamefont{H.}~\bibnamefont{Levine}},
  \bibinfo{author}{\bibfnamefont{A.}~\bibnamefont{Keesling}},
  \bibinfo{author}{\bibfnamefont{G.}~\bibnamefont{Semeghini}},
  \bibinfo{author}{\bibfnamefont{A.}~\bibnamefont{Omran}},
  \bibinfo{author}{\bibfnamefont{T.~T.} \bibnamefont{Wang}},
  \bibinfo{author}{\bibfnamefont{S.}~\bibnamefont{Ebadi}},
  \bibinfo{author}{\bibfnamefont{H.}~\bibnamefont{Bernien}},
  \bibinfo{author}{\bibfnamefont{M.}~\bibnamefont{Greiner}},
  \bibinfo{author}{\bibfnamefont{V.}~\bibnamefont{Vuleti{\'c}}},
  \bibinfo{author}{\bibfnamefont{H.}~\bibnamefont{Pichler}},
  \bibnamefont{et~al.}, \bibinfo{journal}{Phys. Rev. Lett.}
  \textbf{\bibinfo{volume}{123}}, \bibinfo{pages}{170503}
  (\bibinfo{year}{2019}).

\bibitem[{\citenamefont{Young et~al.}(2020)\citenamefont{Young, Eckner, Milner,
  Kedar, Norcia, Oelker, Schine, Ye, and Kaufman}}]{cold:youn20}
\bibinfo{author}{\bibfnamefont{A.~W.} \bibnamefont{Young}},
  \bibinfo{author}{\bibfnamefont{W.~J.} \bibnamefont{Eckner}},
  \bibinfo{author}{\bibfnamefont{W.~R.} \bibnamefont{Milner}},
  \bibinfo{author}{\bibfnamefont{D.}~\bibnamefont{Kedar}},
  \bibinfo{author}{\bibfnamefont{M.~A.} \bibnamefont{Norcia}},
  \bibinfo{author}{\bibfnamefont{E.}~\bibnamefont{Oelker}},
  \bibinfo{author}{\bibfnamefont{N.}~\bibnamefont{Schine}},
  \bibinfo{author}{\bibfnamefont{J.}~\bibnamefont{Ye}}, \bibnamefont{and}
  \bibinfo{author}{\bibfnamefont{A.~M.} \bibnamefont{Kaufman}},
  \bibinfo{journal}{Nat. Phys.} \textbf{\bibinfo{volume}{588}},
  \bibinfo{pages}{408} (\bibinfo{year}{2020}).

\bibitem[{\citenamefont{Kaufman and Ni}(2021)}]{cold:kauf21}
\bibinfo{author}{\bibfnamefont{A.~M.} \bibnamefont{Kaufman}} \bibnamefont{and}
  \bibinfo{author}{\bibfnamefont{K.-K.} \bibnamefont{Ni}},
  \bibinfo{journal}{Nat. Phys.} \textbf{\bibinfo{volume}{17}},
  \bibinfo{pages}{1324} (\bibinfo{year}{2021}).

\bibitem[{\citenamefont{Yu et~al.}(2021)\citenamefont{Yu, Wang, Hood, Picard,
  Zhang, Cairncross, Hutson, Gonzalez-Ferez, Rosenband, and Ni}}]{cold:yicha21}
\bibinfo{author}{\bibfnamefont{Y.}~\bibnamefont{Yu}},
  \bibinfo{author}{\bibfnamefont{K.}~\bibnamefont{Wang}},
  \bibinfo{author}{\bibfnamefont{J.~D.} \bibnamefont{Hood}},
  \bibinfo{author}{\bibfnamefont{L.~R.~B.} \bibnamefont{Picard}},
  \bibinfo{author}{\bibfnamefont{J.~T.} \bibnamefont{Zhang}},
  \bibinfo{author}{\bibfnamefont{W.~B.} \bibnamefont{Cairncross}},
  \bibinfo{author}{\bibfnamefont{J.~M.} \bibnamefont{Hutson}},
  \bibinfo{author}{\bibfnamefont{R.}~\bibnamefont{Gonzalez-Ferez}},
  \bibinfo{author}{\bibfnamefont{T.}~\bibnamefont{Rosenband}},
  \bibnamefont{and} \bibinfo{author}{\bibfnamefont{K.-K.} \bibnamefont{Ni}},
  \bibinfo{journal}{Phys. Rev. X} \textbf{\bibinfo{volume}{11}},
  \bibinfo{pages}{031061} (\bibinfo{year}{2021}).

\bibitem[{\citenamefont{Ti et~al.}(2020)\citenamefont{Ti, Shen, Ho~Thanh, Wen,
  and Liu}}]{cold:ti20}
\bibinfo{author}{\bibfnamefont{C.}~\bibnamefont{Ti}},
  \bibinfo{author}{\bibfnamefont{Y.}~\bibnamefont{Shen}},
  \bibinfo{author}{\bibfnamefont{M.-T.} \bibnamefont{Ho~Thanh}},
  \bibinfo{author}{\bibfnamefont{Q.}~\bibnamefont{Wen}}, \bibnamefont{and}
  \bibinfo{author}{\bibfnamefont{Y.}~\bibnamefont{Liu}}, \bibinfo{journal}{Sci.
  Rep.} \textbf{\bibinfo{volume}{10}}, \bibinfo{pages}{20099}
  (\bibinfo{year}{2020}).

\bibitem[{\citenamefont{Ruttley et~al.}(2023)\citenamefont{Ruttley, Guttridge,
  Spence, Bird, Le~Sueur, Hutson, and Cornish}}]{cold:rutt23}
\bibinfo{author}{\bibfnamefont{D.~K.} \bibnamefont{Ruttley}},
  \bibinfo{author}{\bibfnamefont{A.}~\bibnamefont{Guttridge}},
  \bibinfo{author}{\bibfnamefont{S.}~\bibnamefont{Spence}},
  \bibinfo{author}{\bibfnamefont{R.~C.} \bibnamefont{Bird}},
  \bibinfo{author}{\bibfnamefont{C.~R.} \bibnamefont{Le~Sueur}},
  \bibinfo{author}{\bibfnamefont{J.~M.} \bibnamefont{Hutson}},
  \bibnamefont{and} \bibinfo{author}{\bibfnamefont{S.~L.}
  \bibnamefont{Cornish}}, \bibinfo{journal}{Phys. Rev. Lett.}
  \textbf{\bibinfo{volume}{130}}, \bibinfo{pages}{223401}
  (\bibinfo{year}{2023}).

\bibitem[{\citenamefont{Ruttley et~al.}(2024)\citenamefont{Ruttley, Guttridge,
  Hepworth, and Cornish}}]{cold:rutt24}
\bibinfo{author}{\bibfnamefont{D.~K.} \bibnamefont{Ruttley}},
  \bibinfo{author}{\bibfnamefont{A.}~\bibnamefont{Guttridge}},
  \bibinfo{author}{\bibfnamefont{T.~R.} \bibnamefont{Hepworth}},
  \bibnamefont{and} \bibinfo{author}{\bibfnamefont{S.~L.}
  \bibnamefont{Cornish}}, \bibinfo{journal}{PRX Quantum}
  \textbf{\bibinfo{volume}{5}}, \bibinfo{pages}{020333} (\bibinfo{year}{2024}).

\bibitem[{\citenamefont{Chin et~al.}(2010)\citenamefont{Chin, Grimm, Julienne,
  and Tiesinga}}]{cold:chin10}
\bibinfo{author}{\bibfnamefont{C.}~\bibnamefont{Chin}},
  \bibinfo{author}{\bibfnamefont{R.}~\bibnamefont{Grimm}},
  \bibinfo{author}{\bibfnamefont{P.}~\bibnamefont{Julienne}}, \bibnamefont{and}
  \bibinfo{author}{\bibfnamefont{E.}~\bibnamefont{Tiesinga}},
  \bibinfo{journal}{Rev. Mod. Phys.} \textbf{\bibinfo{volume}{82}},
  \bibinfo{pages}{1225} (\bibinfo{year}{2010}).

\bibitem[{\citenamefont{Aymar}(1978)}]{aies:ayma78}
\bibinfo{author}{\bibfnamefont{M.}~\bibnamefont{Aymar}}, \bibinfo{journal}{J.
  Phys. B} \textbf{\bibinfo{volume}{11}}, \bibinfo{pages}{1413}
  (\bibinfo{year}{1978}).

\bibitem[{\citenamefont{Schneider et~al.}(2009)\citenamefont{Schneider,
  Grishkevich, and Saenz}}]{cold:schn09}
\bibinfo{author}{\bibfnamefont{P.-I.} \bibnamefont{Schneider}},
  \bibinfo{author}{\bibfnamefont{S.}~\bibnamefont{Grishkevich}},
  \bibnamefont{and} \bibinfo{author}{\bibfnamefont{A.}~\bibnamefont{Saenz}},
  \bibinfo{journal}{Phys. Rev. A} \textbf{\bibinfo{volume}{80}},
  \bibinfo{pages}{013404} (\bibinfo{year}{2009}).

\bibitem[{\citenamefont{Sala et~al.}(2012)\citenamefont{Sala, Schneider, and
  Saenz}}]{cold:sala12}
\bibinfo{author}{\bibfnamefont{S.}~\bibnamefont{Sala}},
  \bibinfo{author}{\bibfnamefont{P.-I.} \bibnamefont{Schneider}},
  \bibnamefont{and} \bibinfo{author}{\bibfnamefont{A.}~\bibnamefont{Saenz}},
  \bibinfo{journal}{Phys. Rev. Lett.} \textbf{\bibinfo{volume}{109}},
  \bibinfo{pages}{073201} (\bibinfo{year}{2012}).

\bibitem[{\citenamefont{Haller et~al.}(2010)\citenamefont{Haller, Mark, Hart,
  Danzl, Reich{\-}s{\"o}l{\-}lner, Melezhik, Schmelcher, and
  N{\"a}gerl}}]{cold:hall10b}
\bibinfo{author}{\bibfnamefont{E.}~\bibnamefont{Haller}},
  \bibinfo{author}{\bibfnamefont{M.~J.} \bibnamefont{Mark}},
  \bibinfo{author}{\bibfnamefont{R.}~\bibnamefont{Hart}},
  \bibinfo{author}{\bibfnamefont{J.~G.} \bibnamefont{Danzl}},
  \bibinfo{author}{\bibfnamefont{L.}~\bibnamefont{Reich{\-}s{\"o}l{\-}lner}},
  \bibinfo{author}{\bibfnamefont{V.}~\bibnamefont{Melezhik}},
  \bibinfo{author}{\bibfnamefont{P.}~\bibnamefont{Schmelcher}},
  \bibnamefont{and} \bibinfo{author}{\bibfnamefont{H.-C.}
  \bibnamefont{N{\"a}gerl}}, \bibinfo{journal}{Phys. Rev. Lett.}
  \textbf{\bibinfo{volume}{104}}, \bibinfo{pages}{153203}
  (\bibinfo{year}{2010}).

\bibitem[{\citenamefont{Capecchi et~al.}(2023)\citenamefont{Capecchi,
  Cantillano, Mark, Meinert, Schindewolf, Landini, Saenz, Revuelta, and
  N{\"a}gerl}}]{cold:cape23}
\bibinfo{author}{\bibfnamefont{D.}~\bibnamefont{Capecchi}},
  \bibinfo{author}{\bibfnamefont{C.}~\bibnamefont{Cantillano}},
  \bibinfo{author}{\bibfnamefont{M.~J.} \bibnamefont{Mark}},
  \bibinfo{author}{\bibfnamefont{F.}~\bibnamefont{Meinert}},
  \bibinfo{author}{\bibfnamefont{A.}~\bibnamefont{Schindewolf}},
  \bibinfo{author}{\bibfnamefont{M.}~\bibnamefont{Landini}},
  \bibinfo{author}{\bibfnamefont{A.}~\bibnamefont{Saenz}},
  \bibinfo{author}{\bibfnamefont{F.}~\bibnamefont{Revuelta}}, \bibnamefont{and}
  \bibinfo{author}{\bibfnamefont{H.-C.} \bibnamefont{N{\"a}gerl}},
  \bibinfo{journal}{Phys. Rev. Lett.} \textbf{\bibinfo{volume}{131}},
  \bibinfo{pages}{213002} (\bibinfo{year}{2023}).

\bibitem[{\citenamefont{Valiente and M{\o}lmer}(2011)}]{cold:vali11}
\bibinfo{author}{\bibfnamefont{M.}~\bibnamefont{Valiente}} \bibnamefont{and}
  \bibinfo{author}{\bibfnamefont{K.}~\bibnamefont{M{\o}lmer}},
  \bibinfo{journal}{Phys. Rev. A} \textbf{\bibinfo{volume}{84}},
  \bibinfo{pages}{053628} (\bibinfo{year}{2011}).

\bibitem[{\citenamefont{Melezhik and Schmelcher}(2011)}]{cold:mele11}
\bibinfo{author}{\bibfnamefont{V.~S.} \bibnamefont{Melezhik}} \bibnamefont{and}
  \bibinfo{author}{\bibfnamefont{P.}~\bibnamefont{Schmelcher}},
  \bibinfo{journal}{Phys. Rev. A} \textbf{\bibinfo{volume}{84}},
  \bibinfo{pages}{042712} (\bibinfo{year}{2011}).

\bibitem[{\citenamefont{Sala et~al.}(2013)\citenamefont{Sala, Z{\"u}rn, Lompe,
  Wenz, Murmann, Serwane, Jochim, and Saenz}}]{cold:sala13}
\bibinfo{author}{\bibfnamefont{S.}~\bibnamefont{Sala}},
  \bibinfo{author}{\bibfnamefont{G.}~\bibnamefont{Z{\"u}rn}},
  \bibinfo{author}{\bibfnamefont{T.}~\bibnamefont{Lompe}},
  \bibinfo{author}{\bibfnamefont{A.}~\bibnamefont{Wenz}},
  \bibinfo{author}{\bibfnamefont{S.}~\bibnamefont{Murmann}},
  \bibinfo{author}{\bibfnamefont{F.}~\bibnamefont{Serwane}},
  \bibinfo{author}{\bibfnamefont{S.}~\bibnamefont{Jochim}}, \bibnamefont{and}
  \bibinfo{author}{\bibfnamefont{A.}~\bibnamefont{Saenz}},
  \bibinfo{journal}{Phys. Rev. Lett.} \textbf{\bibinfo{volume}{110}},
  \bibinfo{pages}{203202} (\bibinfo{year}{2013}).

\bibitem[{\citenamefont{Sala and Saenz}(2016)}]{cold:sala16a}
\bibinfo{author}{\bibfnamefont{S.}~\bibnamefont{Sala}} \bibnamefont{and}
  \bibinfo{author}{\bibfnamefont{A.}~\bibnamefont{Saenz}},
  \bibinfo{journal}{Phys. Rev. A} \textbf{\bibinfo{volume}{94}},
  \bibinfo{pages}{022713} (\bibinfo{year}{2016}).

\bibitem[{\citenamefont{Efimov}(1970{\natexlab{a}})}]{Efimov1970}
\bibinfo{author}{\bibfnamefont{V.~N.} \bibnamefont{Efimov}},
  \bibinfo{journal}{Yad. Fiz.} \textbf{\bibinfo{volume}{12}},
  \bibinfo{pages}{1080} (\bibinfo{year}{1970}{\natexlab{a}}).

\bibitem[{\citenamefont{Efimov}(1970{\natexlab{b}})}]{Efimov1970a}
\bibinfo{author}{\bibfnamefont{V.~N.} \bibnamefont{Efimov}},
  \bibinfo{journal}{Phys. Lett. B} \textbf{\bibinfo{volume}{33}},
  \bibinfo{pages}{563} (\bibinfo{year}{1970}{\natexlab{b}}).

\bibitem[{\citenamefont{Esry et~al.}(1999)\citenamefont{Esry, Greene, and
  Burke}}]{cold:esry99a}
\bibinfo{author}{\bibfnamefont{B.~D.} \bibnamefont{Esry}},
  \bibinfo{author}{\bibfnamefont{C.~H.} \bibnamefont{Greene}},
  \bibnamefont{and} \bibinfo{author}{\bibfnamefont{J.~P.} \bibnamefont{Burke}},
  \bibinfo{journal}{Phys. Rev. Lett.} \textbf{\bibinfo{volume}{83}},
  \bibinfo{pages}{1751} (\bibinfo{year}{1999}).

\bibitem[{\citenamefont{Werner and Castin}(2006)}]{cold:wern06}
\bibinfo{author}{\bibfnamefont{F.}~\bibnamefont{Werner}} \bibnamefont{and}
  \bibinfo{author}{\bibfnamefont{Y.}~\bibnamefont{Castin}},
  \bibinfo{journal}{Phys. Rev. Lett.} \textbf{\bibinfo{volume}{97}},
  \bibinfo{pages}{150401} (\bibinfo{year}{2006}).

\bibitem[{\citenamefont{Ferlaino et~al.}(2009)\citenamefont{Ferlaino, Knoop,
  Berninger, Harm, D'Incao, N\"agerl, and Grimm}}]{cold:ferl09}
\bibinfo{author}{\bibfnamefont{F.}~\bibnamefont{Ferlaino}},
  \bibinfo{author}{\bibfnamefont{S.}~\bibnamefont{Knoop}},
  \bibinfo{author}{\bibfnamefont{M.}~\bibnamefont{Berninger}},
  \bibinfo{author}{\bibfnamefont{W.}~\bibnamefont{Harm}},
  \bibinfo{author}{\bibfnamefont{J.~P.} \bibnamefont{D'Incao}},
  \bibinfo{author}{\bibfnamefont{H.-C.} \bibnamefont{N\"agerl}},
  \bibnamefont{and} \bibinfo{author}{\bibfnamefont{R.}~\bibnamefont{Grimm}},
  \bibinfo{journal}{Phys. Rev. Lett.} \textbf{\bibinfo{volume}{102}},
  \bibinfo{pages}{140401} (\bibinfo{year}{2009}).

\bibitem[{\citenamefont{Naidon and Endo}(2017)}]{cold:naid17}
\bibinfo{author}{\bibfnamefont{P.}~\bibnamefont{Naidon}} \bibnamefont{and}
  \bibinfo{author}{\bibfnamefont{S.}~\bibnamefont{Endo}},
  \bibinfo{journal}{Rep. Prog. Phys.} \textbf{\bibinfo{volume}{80}},
  \bibinfo{pages}{056001} (\bibinfo{year}{2017}).

\bibitem[{\citenamefont{Higgins and Greene}(2022)}]{cold:higg22}
\bibinfo{author}{\bibfnamefont{M.~D.} \bibnamefont{Higgins}} \bibnamefont{and}
  \bibinfo{author}{\bibfnamefont{C.~H.} \bibnamefont{Greene}},
  \bibinfo{journal}{Phys. Rev. A} \textbf{\bibinfo{volume}{106}},
  \bibinfo{pages}{023304} (\bibinfo{year}{2022}).

\bibitem[{\citenamefont{Bougas et~al.}(2023)\citenamefont{Bougas, Mistakidis,
  Schmelcher, Greene, and Giannakeas}}]{cold:boug23}
\bibinfo{author}{\bibfnamefont{G.}~\bibnamefont{Bougas}},
  \bibinfo{author}{\bibfnamefont{S.}~\bibnamefont{Mistakidis}},
  \bibinfo{author}{\bibfnamefont{P.}~\bibnamefont{Schmelcher}},
  \bibinfo{author}{\bibfnamefont{C.}~\bibnamefont{Greene}}, \bibnamefont{and}
  \bibinfo{author}{\bibfnamefont{P.}~\bibnamefont{Giannakeas}},
  \bibinfo{journal}{Phys. Rev. Res.} \textbf{\bibinfo{volume}{5}},
  \bibinfo{pages}{043134} (\bibinfo{year}{2023}).

\bibitem[{\citenamefont{Grishkevich and Saenz}(2009)}]{cold:gris09}
\bibinfo{author}{\bibfnamefont{S.}~\bibnamefont{Grishkevich}} \bibnamefont{and}
  \bibinfo{author}{\bibfnamefont{A.}~\bibnamefont{Saenz}},
  \bibinfo{journal}{Phys. Rev. A} \textbf{\bibinfo{volume}{80}},
  \bibinfo{pages}{013403} (\bibinfo{year}{2009}).

\bibitem[{\citenamefont{Bougas et~al.}(2022)\citenamefont{Bougas, Mistakidis,
  Giannakeas, and Schmelcher}}]{cold:boug22}
\bibinfo{author}{\bibfnamefont{G.}~\bibnamefont{Bougas}},
  \bibinfo{author}{\bibfnamefont{S.}~\bibnamefont{Mistakidis}},
  \bibinfo{author}{\bibfnamefont{P.}~\bibnamefont{Giannakeas}},
  \bibnamefont{and}
  \bibinfo{author}{\bibfnamefont{P.}~\bibnamefont{Schmelcher}},
  \bibinfo{journal}{Phys. Rev. A} \textbf{\bibinfo{volume}{106}},
  \bibinfo{pages}{043323} (\bibinfo{year}{2022}).

\bibitem[{\citenamefont{Bougas et~al.}(2019)\citenamefont{Bougas, Mistakidis,
  and Schmelcher}}]{cold:boug19}
\bibinfo{author}{\bibfnamefont{G.}~\bibnamefont{Bougas}},
  \bibinfo{author}{\bibfnamefont{S.}~\bibnamefont{Mistakidis}},
  \bibnamefont{and}
  \bibinfo{author}{\bibfnamefont{P.}~\bibnamefont{Schmelcher}},
  \bibinfo{journal}{Phys. Rev. A} \textbf{\bibinfo{volume}{100}},
  \bibinfo{pages}{053602} (\bibinfo{year}{2019}).

\bibitem[{\citenamefont{Bougas et~al.}(2021)\citenamefont{Bougas, Mistakidis,
  Giannakeas, and Schmelcher}}]{cold:boug21}
\bibinfo{author}{\bibfnamefont{G.}~\bibnamefont{Bougas}},
  \bibinfo{author}{\bibfnamefont{S.}~\bibnamefont{Mistakidis}},
  \bibinfo{author}{\bibfnamefont{P.}~\bibnamefont{Giannakeas}},
  \bibnamefont{and}
  \bibinfo{author}{\bibfnamefont{P.}~\bibnamefont{Schmelcher}},
  \bibinfo{journal}{New J. Phys.} \textbf{\bibinfo{volume}{23}},
  \bibinfo{pages}{093022} (\bibinfo{year}{2021}).

\bibitem[{\citenamefont{Esry and Greene}(1999)}]{cold:esry99}
\bibinfo{author}{\bibfnamefont{B.~D.} \bibnamefont{Esry}} \bibnamefont{and}
  \bibinfo{author}{\bibfnamefont{C.~H.} \bibnamefont{Greene}},
  \bibinfo{journal}{Phys. Rev. A} \textbf{\bibinfo{volume}{60}},
  \bibinfo{pages}{1451} (\bibinfo{year}{1999}).

\bibitem[{\citenamefont{Brauneis et~al.}(2025)\citenamefont{Brauneis, Hammer,
  Reimann, and Volosniev}}]{cold:brau25}
\bibinfo{author}{\bibfnamefont{F.}~\bibnamefont{Brauneis}},
  \bibinfo{author}{\bibfnamefont{H.-W.} \bibnamefont{Hammer}},
  \bibinfo{author}{\bibfnamefont{S.~M.} \bibnamefont{Reimann}},
  \bibnamefont{and} \bibinfo{author}{\bibfnamefont{A.~G.}
  \bibnamefont{Volosniev}}, \bibinfo{journal}{Phys. Rev. A}
  \textbf{\bibinfo{volume}{111}}, \bibinfo{pages}{013303}
  (\bibinfo{year}{2025}).

\bibitem[{\citenamefont{Blume and Greene}(2002)}]{cold:blum02}
\bibinfo{author}{\bibfnamefont{D.}~\bibnamefont{Blume}} \bibnamefont{and}
  \bibinfo{author}{\bibfnamefont{C.~H.} \bibnamefont{Greene}},
  \bibinfo{journal}{Phys. Rev. A} \textbf{\bibinfo{volume}{65}},
  \bibinfo{pages}{043613} (\bibinfo{year}{2002}).

\bibitem[{\citenamefont{Armstrong et~al.}(2011)\citenamefont{Armstrong, Zinner,
  Fedorov, and Jensen}}]{cold:arms11}
\bibinfo{author}{\bibfnamefont{J.~R.} \bibnamefont{Armstrong}},
  \bibinfo{author}{\bibfnamefont{N.~T.} \bibnamefont{Zinner}},
  \bibinfo{author}{\bibfnamefont{D.~V.} \bibnamefont{Fedorov}},
  \bibnamefont{and} \bibinfo{author}{\bibfnamefont{A.~S.}
  \bibnamefont{Jensen}}, \bibinfo{journal}{J. Phys. B}
  \textbf{\bibinfo{volume}{44}}, \bibinfo{pages}{055303}
  (\bibinfo{year}{2011}).

\bibitem[{\citenamefont{Yin et~al.}(2014)\citenamefont{Yin, Blume, Johnson, and
  Tiesinga}}]{cold:yin14}
\bibinfo{author}{\bibfnamefont{X.}~\bibnamefont{Yin}},
  \bibinfo{author}{\bibfnamefont{D.}~\bibnamefont{Blume}},
  \bibinfo{author}{\bibfnamefont{P.}~\bibnamefont{Johnson}}, \bibnamefont{and}
  \bibinfo{author}{\bibfnamefont{E.}~\bibnamefont{Tiesinga}},
  \bibinfo{journal}{Phys. Rev. A} \textbf{\bibinfo{volume}{90}},
  \bibinfo{pages}{043631} (\bibinfo{year}{2014}).

\bibitem[{\citenamefont{Jeszenszki et~al.}(2018)\citenamefont{Jeszenszki,
  Cherny, and Brand}}]{cold:jesz18}
\bibinfo{author}{\bibfnamefont{P.}~\bibnamefont{Jeszenszki}},
  \bibinfo{author}{\bibfnamefont{A.~Y.} \bibnamefont{Cherny}},
  \bibnamefont{and} \bibinfo{author}{\bibfnamefont{J.}~\bibnamefont{Brand}},
  \bibinfo{journal}{Phys. Rev. A} \textbf{\bibinfo{volume}{97}},
  \bibinfo{pages}{042708} (\bibinfo{year}{2018}).

\bibitem[{\citenamefont{Jeszenszki et~al.}(2019)\citenamefont{Jeszenszki,
  Alavi, and Brand}}]{cold:jesz19}
\bibinfo{author}{\bibfnamefont{P.}~\bibnamefont{Jeszenszki}},
  \bibinfo{author}{\bibfnamefont{A.}~\bibnamefont{Alavi}}, \bibnamefont{and}
  \bibinfo{author}{\bibfnamefont{J.}~\bibnamefont{Brand}},
  \bibinfo{journal}{Phys. Rev. A} \textbf{\bibinfo{volume}{99}},
  \bibinfo{pages}{033608} (\bibinfo{year}{2019}).

\bibitem[{\citenamefont{Boys}(1950)}]{math:boys50}
\bibinfo{author}{\bibfnamefont{S.~F.} \bibnamefont{Boys}},
  \bibinfo{journal}{Proc. R. Soc. Lond. A} \textbf{\bibinfo{volume}{200}},
  \bibinfo{pages}{542} (\bibinfo{year}{1950}).

\bibitem[{\citenamefont{Rakshit and Blume}(2012)}]{cold:raks12}
\bibinfo{author}{\bibfnamefont{D.}~\bibnamefont{Rakshit}} \bibnamefont{and}
  \bibinfo{author}{\bibfnamefont{D.}~\bibnamefont{Blume}},
  \bibinfo{journal}{Phys. Rev. A} \textbf{\bibinfo{volume}{86}},
  \bibinfo{pages}{062513} (\bibinfo{year}{2012}).

\bibitem[{\citenamefont{Barca and Loos}(2017)}]{math:barc17}
\bibinfo{author}{\bibfnamefont{G.~M.} \bibnamefont{Barca}} \bibnamefont{and}
  \bibinfo{author}{\bibfnamefont{P.-F.} \bibnamefont{Loos}},
  \bibinfo{journal}{J. Chem. Phys.} \textbf{\bibinfo{volume}{147}},
  \bibinfo{pages}{2} (\bibinfo{year}{2017}).

\bibitem[{\citenamefont{Grishkevich et~al.}(2011)\citenamefont{Grishkevich,
  Sala, and Saenz}}]{cold:gris11}
\bibinfo{author}{\bibfnamefont{S.}~\bibnamefont{Grishkevich}},
  \bibinfo{author}{\bibfnamefont{S.}~\bibnamefont{Sala}}, \bibnamefont{and}
  \bibinfo{author}{\bibfnamefont{A.}~\bibnamefont{Saenz}},
  \bibinfo{journal}{Phys. Rev. A} \textbf{\bibinfo{volume}{84}},
  \bibinfo{pages}{062710} (\bibinfo{year}{2011}).

\bibitem[{\citenamefont{Schulz et~al.}(2015)\citenamefont{Schulz, Sala, and
  Saenz}}]{cold:schu15}
\bibinfo{author}{\bibfnamefont{B.}~\bibnamefont{Schulz}},
  \bibinfo{author}{\bibfnamefont{S.}~\bibnamefont{Sala}}, \bibnamefont{and}
  \bibinfo{author}{\bibfnamefont{A.}~\bibnamefont{Saenz}},
  \bibinfo{journal}{New J. Phys.} \textbf{\bibinfo{volume}{17}},
  \bibinfo{pages}{065002} (\bibinfo{year}{2015}).

\bibitem[{\citenamefont{Schulz and Saenz}(2016)}]{cold:schu16}
\bibinfo{author}{\bibfnamefont{B.}~\bibnamefont{Schulz}} \bibnamefont{and}
  \bibinfo{author}{\bibfnamefont{A.}~\bibnamefont{Saenz}},
  \bibinfo{journal}{Comp. Phys. Comm.} \textbf{\bibinfo{volume}{17}},
  \bibinfo{pages}{3747} (\bibinfo{year}{2016}).

\bibitem[{\citenamefont{Daily and Blume}(2010)}]{cold:dail10}
\bibinfo{author}{\bibfnamefont{K.~M.} \bibnamefont{Daily}} \bibnamefont{and}
  \bibinfo{author}{\bibfnamefont{D.}~\bibnamefont{Blume}},
  \bibinfo{journal}{Phys. Rev. A} \textbf{\bibinfo{volume}{81}},
  \bibinfo{pages}{053615} (\bibinfo{year}{2010}).

\bibitem[{\citenamefont{Rotureau et~al.}(2010)\citenamefont{Rotureau, Stetcu,
  Barrett, Birse, and van Kolck}}]{cold:rotu10}
\bibinfo{author}{\bibfnamefont{J.}~\bibnamefont{Rotureau}},
  \bibinfo{author}{\bibfnamefont{I.}~\bibnamefont{Stetcu}},
  \bibinfo{author}{\bibfnamefont{B.~R.} \bibnamefont{Barrett}},
  \bibinfo{author}{\bibfnamefont{M.~C.} \bibnamefont{Birse}}, \bibnamefont{and}
  \bibinfo{author}{\bibfnamefont{U.}~\bibnamefont{van Kolck}},
  \bibinfo{journal}{Phys. Rev. A} \textbf{\bibinfo{volume}{82}},
  \bibinfo{pages}{032711} (\bibinfo{year}{2010}).

\bibitem[{\citenamefont{Greene et~al.}(2017)\citenamefont{Greene, Giannakeas,
  and P{\'e}rez-R{\'\i}os}}]{cold:gree17}
\bibinfo{author}{\bibfnamefont{C.~H.} \bibnamefont{Greene}},
  \bibinfo{author}{\bibfnamefont{P.}~\bibnamefont{Giannakeas}},
  \bibnamefont{and}
  \bibinfo{author}{\bibfnamefont{J.}~\bibnamefont{P{\'e}rez-R{\'\i}os}},
  \bibinfo{journal}{Rev. Mod. Phys.} \textbf{\bibinfo{volume}{89}},
  \bibinfo{pages}{035006} (\bibinfo{year}{2017}).

\bibitem[{\citenamefont{Hiyama and Kamimura}(2018)}]{gen:hiya18}
\bibinfo{author}{\bibfnamefont{E.}~\bibnamefont{Hiyama}} \bibnamefont{and}
  \bibinfo{author}{\bibfnamefont{M.}~\bibnamefont{Kamimura}},
  \bibinfo{journal}{Front. Phys.} \textbf{\bibinfo{volume}{13}},
  \bibinfo{pages}{132106} (\bibinfo{year}{2018}).

\bibitem[{\citenamefont{von Stecher and Greene}(2009)}]{cold:von09}
\bibinfo{author}{\bibfnamefont{J.}~\bibnamefont{von Stecher}} \bibnamefont{and}
  \bibinfo{author}{\bibfnamefont{C.~H.} \bibnamefont{Greene}},
  \bibinfo{journal}{Phys. Rev. A} \textbf{\bibinfo{volume}{80}},
  \bibinfo{pages}{022504} (\bibinfo{year}{2009}).

\bibitem[{\citenamefont{Sze et~al.}(2018)\citenamefont{Sze, Sykes, Blume, and
  Bohn}}]{cold:sze18}
\bibinfo{author}{\bibfnamefont{M.}~\bibnamefont{Sze}},
  \bibinfo{author}{\bibfnamefont{A.}~\bibnamefont{Sykes}},
  \bibinfo{author}{\bibfnamefont{D.}~\bibnamefont{Blume}}, \bibnamefont{and}
  \bibinfo{author}{\bibfnamefont{J.}~\bibnamefont{Bohn}},
  \bibinfo{journal}{Phys. Rev. A} \textbf{\bibinfo{volume}{97}},
  \bibinfo{pages}{033608} (\bibinfo{year}{2018}).

\bibitem[{\citenamefont{Sala et~al.}(2017)\citenamefont{Sala, F{\"o}rster, and
  Saenz}}]{cold:sala17}
\bibinfo{author}{\bibfnamefont{S.}~\bibnamefont{Sala}},
  \bibinfo{author}{\bibfnamefont{J.}~\bibnamefont{F{\"o}rster}},
  \bibnamefont{and} \bibinfo{author}{\bibfnamefont{A.}~\bibnamefont{Saenz}},
  \bibinfo{journal}{Phys. Rev. A} \textbf{\bibinfo{volume}{95}},
  \bibinfo{pages}{011403(R)} (\bibinfo{year}{2017}).

\bibitem[{\citenamefont{Silkowski et~al.}(2015)\citenamefont{Silkowski, Lesiuk,
  and Moszynski}}]{math:silk15a}
\bibinfo{author}{\bibfnamefont{M.}~\bibnamefont{Silkowski}},
  \bibinfo{author}{\bibfnamefont{M.}~\bibnamefont{Lesiuk}}, \bibnamefont{and}
  \bibinfo{author}{\bibfnamefont{R.}~\bibnamefont{Moszynski}},
  \bibinfo{journal}{J. Chem. Phys.} \textbf{\bibinfo{volume}{142}},
  \bibinfo{pages}{124102} (\bibinfo{year}{2015}).

\bibitem[{\citenamefont{Balcerzak et~al.}(2017)\citenamefont{Balcerzak, Lesiuk,
  and Moszynski}}]{gen:balc17}
\bibinfo{author}{\bibfnamefont{J.~G.} \bibnamefont{Balcerzak}},
  \bibinfo{author}{\bibfnamefont{M.}~\bibnamefont{Lesiuk}}, \bibnamefont{and}
  \bibinfo{author}{\bibfnamefont{R.}~\bibnamefont{Moszynski}},
  \bibinfo{journal}{Phys. Rev. A} \textbf{\bibinfo{volume}{96}},
  \bibinfo{pages}{052510} (\bibinfo{year}{2017}).

\bibitem[{\citenamefont{Deutsch}(1991)}]{cold:deut91}
\bibinfo{author}{\bibfnamefont{J.~M.} \bibnamefont{Deutsch}},
  \bibinfo{journal}{Phys. Rev. A} \textbf{\bibinfo{volume}{43}},
  \bibinfo{pages}{2046} (\bibinfo{year}{1991}).

\bibitem[{\citenamefont{Srednicki}(1994)}]{cold:sred94}
\bibinfo{author}{\bibfnamefont{M.}~\bibnamefont{Srednicki}},
  \bibinfo{journal}{Phys. Rev. E} \textbf{\bibinfo{volume}{50}},
  \bibinfo{pages}{888} (\bibinfo{year}{1994}).

\bibitem[{\citenamefont{Shiraishi and Matsumoto}(2021)}]{cold:shir21}
\bibinfo{author}{\bibfnamefont{N.}~\bibnamefont{Shiraishi}} \bibnamefont{and}
  \bibinfo{author}{\bibfnamefont{K.}~\bibnamefont{Matsumoto}},
  \bibinfo{journal}{Nat. Commun.} \textbf{\bibinfo{volume}{12}},
  \bibinfo{pages}{5084} (\bibinfo{year}{2021}).

\bibitem[{\citenamefont{Moudgalya et~al.}(2018)\citenamefont{Moudgalya,
  Regnault, and Bernevig}}]{cold:moud18}
\bibinfo{author}{\bibfnamefont{S.}~\bibnamefont{Moudgalya}},
  \bibinfo{author}{\bibfnamefont{N.}~\bibnamefont{Regnault}}, \bibnamefont{and}
  \bibinfo{author}{\bibfnamefont{B.~A.} \bibnamefont{Bernevig}},
  \bibinfo{journal}{Phys. Rev. B} \textbf{\bibinfo{volume}{98}},
  \bibinfo{pages}{235156} (\bibinfo{year}{2018}).

\bibitem[{\citenamefont{Turner et~al.}(2018)\citenamefont{Turner, Michailidis,
  Abanin, Serbyn, and Papi{\'c}}}]{cold:turn18}
\bibinfo{author}{\bibfnamefont{C.~J.} \bibnamefont{Turner}},
  \bibinfo{author}{\bibfnamefont{A.~A.} \bibnamefont{Michailidis}},
  \bibinfo{author}{\bibfnamefont{D.~A.} \bibnamefont{Abanin}},
  \bibinfo{author}{\bibfnamefont{M.}~\bibnamefont{Serbyn}}, \bibnamefont{and}
  \bibinfo{author}{\bibfnamefont{Z.}~\bibnamefont{Papi{\'c}}},
  \bibinfo{journal}{Nat. Phys.} \textbf{\bibinfo{volume}{14}},
  \bibinfo{pages}{745} (\bibinfo{year}{2018}).

\bibitem[{\citenamefont{Serbyn et~al.}(2021)\citenamefont{Serbyn, Abanin, and
  Papi{\'c}}}]{cold:serb21}
\bibinfo{author}{\bibfnamefont{M.}~\bibnamefont{Serbyn}},
  \bibinfo{author}{\bibfnamefont{D.~A.} \bibnamefont{Abanin}},
  \bibnamefont{and}
  \bibinfo{author}{\bibfnamefont{Z.}~\bibnamefont{Papi{\'c}}},
  \bibinfo{journal}{Nat. Phys.} \textbf{\bibinfo{volume}{17}},
  \bibinfo{pages}{675} (\bibinfo{year}{2021}).

\bibitem[{\citenamefont{Revuelta et~al.}(2012)\citenamefont{Revuelta, Vergini,
  Benito, and Borondo}}]{cold:revu12}
\bibinfo{author}{\bibfnamefont{F.}~\bibnamefont{Revuelta}},
  \bibinfo{author}{\bibfnamefont{E.~G.} \bibnamefont{Vergini}},
  \bibinfo{author}{\bibfnamefont{R.}~\bibnamefont{Benito}}, \bibnamefont{and}
  \bibinfo{author}{\bibfnamefont{F.}~\bibnamefont{Borondo}},
  \bibinfo{journal}{Phys. Rev. E} \textbf{\bibinfo{volume}{85}},
  \bibinfo{pages}{026214} (\bibinfo{year}{2012}).

\bibitem[{\citenamefont{Szabo and Ostlund}(1996)}]{gen:szab96}
\bibinfo{author}{\bibfnamefont{A.}~\bibnamefont{Szabo}} \bibnamefont{and}
  \bibinfo{author}{\bibfnamefont{N.~S.} \bibnamefont{Ostlund}},
  \emph{\bibinfo{title}{Modern Quantum Chemistry: Introduction to Advanced
  Electronic Structure Theory}} (\bibinfo{publisher}{Courier Corporation},
  \bibinfo{year}{1996}).

\bibitem[{\citenamefont{Schneider et~al.}(2013)\citenamefont{Schneider,
  Grishkevich, and Saenz}}]{cold:schn13}
\bibinfo{author}{\bibfnamefont{P.-I.} \bibnamefont{Schneider}},
  \bibinfo{author}{\bibfnamefont{S.}~\bibnamefont{Grishkevich}},
  \bibnamefont{and} \bibinfo{author}{\bibfnamefont{A.}~\bibnamefont{Saenz}},
  \bibinfo{journal}{Phys. Rev. A} \textbf{\bibinfo{volume}{87}},
  \bibinfo{pages}{053413} (\bibinfo{year}{2013}).

\bibitem[{\citenamefont{Herzberg}(1950)}]{gen:herz50}
\bibinfo{author}{\bibfnamefont{G.}~\bibnamefont{Herzberg}},
  \emph{\bibinfo{title}{Molecular Spectra and Molecular Structure. Vol. 1:
  Spectra of Diatomic Molecules}} (\bibinfo{publisher}{New York: Van Nostrand
  Reinhold}, \bibinfo{year}{1950}).

\bibitem[{\citenamefont{Morse}(1929)}]{gen:mors29}
\bibinfo{author}{\bibfnamefont{P.}~\bibnamefont{Morse}},
  \bibinfo{journal}{Phys. Rev.} \textbf{\bibinfo{volume}{34}},
  \bibinfo{pages}{57} (\bibinfo{year}{1929}).

\bibitem[{\citenamefont{Buckingham}(1938)}]{gen:buck38}
\bibinfo{author}{\bibfnamefont{R.~A.} \bibnamefont{Buckingham}},
  \bibinfo{journal}{Proc. R. Soc. Lond. A} \textbf{\bibinfo{volume}{168}},
  \bibinfo{pages}{264} (\bibinfo{year}{1938}).

\bibitem[{\citenamefont{McMurchie and Davidson}(1978)}]{math:mcmu78}
\bibinfo{author}{\bibfnamefont{L.~E.} \bibnamefont{McMurchie}}
  \bibnamefont{and} \bibinfo{author}{\bibfnamefont{E.~R.}
  \bibnamefont{Davidson}}, \bibinfo{journal}{J. Comp. Phys.}
  \textbf{\bibinfo{volume}{26}}, \bibinfo{pages}{218} (\bibinfo{year}{1978}).

\bibitem[{\citenamefont{Samson et~al.}(2002)\citenamefont{Samson, Klopper, and
  Helgaker}}]{math:sams02}
\bibinfo{author}{\bibfnamefont{C.~C.~M.} \bibnamefont{Samson}},
  \bibinfo{author}{\bibfnamefont{W.}~\bibnamefont{Klopper}}, \bibnamefont{and}
  \bibinfo{author}{\bibfnamefont{T.}~\bibnamefont{Helgaker}},
  \bibinfo{journal}{Comput. Phys. Commun.} \textbf{\bibinfo{volume}{149}},
  \bibinfo{pages}{1} (\bibinfo{year}{2002}).

\bibitem[{\citenamefont{Hobson}(1965)}]{gen:hobs65}
\bibinfo{author}{\bibfnamefont{E.~W.} \bibnamefont{Hobson}},
  \emph{\bibinfo{title}{The Theory of Spherical and Ellipsoidal Harmonics}}
  (\bibinfo{publisher}{Chelsea}, \bibinfo{address}{New York},
  \bibinfo{year}{1965}).

\bibitem[{\citenamefont{Schlegel and Frisch}(1995)}]{math:schl95}
\bibinfo{author}{\bibfnamefont{H.~B.} \bibnamefont{Schlegel}} \bibnamefont{and}
  \bibinfo{author}{\bibfnamefont{M.~J.} \bibnamefont{Frisch}},
  \bibinfo{journal}{Int. J. Quantum Chem.} \textbf{\bibinfo{volume}{54}},
  \bibinfo{pages}{83} (\bibinfo{year}{1995}).

\bibitem[{\citenamefont{Loftus et~al.}(2002)\citenamefont{Loftus, Regal,
  Ticknor, Bohn, and Jin}}]{cold:loft02}
\bibinfo{author}{\bibfnamefont{T.}~\bibnamefont{Loftus}},
  \bibinfo{author}{\bibfnamefont{C.~A.} \bibnamefont{Regal}},
  \bibinfo{author}{\bibfnamefont{C.}~\bibnamefont{Ticknor}},
  \bibinfo{author}{\bibfnamefont{J.~L.} \bibnamefont{Bohn}}, \bibnamefont{and}
  \bibinfo{author}{\bibfnamefont{D.~S.} \bibnamefont{Jin}},
  \bibinfo{journal}{Phys. Rev. Lett.} \textbf{\bibinfo{volume}{88}},
  \bibinfo{pages}{173201} (\bibinfo{year}{2002}).

\bibitem[{\citenamefont{Grishkevich et~al.}(2010)\citenamefont{Grishkevich,
  Schneider, Vanne, and Saenz}}]{cold:gris10}
\bibinfo{author}{\bibfnamefont{S.}~\bibnamefont{Grishkevich}},
  \bibinfo{author}{\bibfnamefont{P.-I.} \bibnamefont{Schneider}},
  \bibinfo{author}{\bibfnamefont{Y.~V.} \bibnamefont{Vanne}}, \bibnamefont{and}
  \bibinfo{author}{\bibfnamefont{A.}~\bibnamefont{Saenz}},
  \bibinfo{journal}{Phys. Rev. A} \textbf{\bibinfo{volume}{81}},
  \bibinfo{pages}{022719} (\bibinfo{year}{2010}).

\bibitem[{\citenamefont{Weiner et~al.}(1999)\citenamefont{Weiner, Bagnato,
  Zilio, and Julienne}}]{cold:wein99}
\bibinfo{author}{\bibfnamefont{J.}~\bibnamefont{Weiner}},
  \bibinfo{author}{\bibfnamefont{V.~S.} \bibnamefont{Bagnato}},
  \bibinfo{author}{\bibfnamefont{S.}~\bibnamefont{Zilio}}, \bibnamefont{and}
  \bibinfo{author}{\bibfnamefont{P.~S.} \bibnamefont{Julienne}},
  \bibinfo{journal}{Rev. Mod. Phys.} \textbf{\bibinfo{volume}{71}},
  \bibinfo{pages}{1} (\bibinfo{year}{1999}).

\bibitem[{\citenamefont{Sala}(2016)}]{sala2016}
\bibinfo{author}{\bibfnamefont{S.}~\bibnamefont{Sala}}, \bibinfo{type}{Phd
  thesis}, \bibinfo{school}{Humboldt-Universität zu Berlin},
  \bibinfo{address}{Germany} (\bibinfo{year}{2016}).

\end{thebibliography}
\end{document}